\documentclass[11pt]{iopart}
\usepackage{comment}
\usepackage{enumerate}
\usepackage{amssymb}
\expandafter\let\csname equation*\endcsname\relax
\expandafter\let\csname endequation*\endcsname\relax
\usepackage{amsmath}
\usepackage{braket}
\usepackage{dsfont}
\usepackage{graphicx}
\usepackage[usenames,dvipsnames]{color}
\usepackage{tensor}
\usepackage{empheq}
\usepackage{capt-of}
\usepackage[normalem]{ulem} 
\usepackage{soul} %
\usepackage{MnSymbol}
\usepackage{cancel}

\usepackage[colorlinks,bookmarks=false,citecolor=NavyBlue,linkcolor=OliveGreen,urlcolor=blue]{hyperref}
\newcommand{\be}{\begin{equation}}
\newcommand{\ee}{\end{equation}}
\newcommand{\ba}{\begin{aligned}}
\newcommand{\ea}{\end{aligned}}
\newcommand{\bw}{\begin{widetext}}
\newcommand{\ew}{\end{widetext}}

\renewcommand{\vec}[1]{\boldsymbol{#1}}
\newcommand{\bea}{\begin{eqnarray}}
\newcommand{\eea}{\end{eqnarray}}

\def\doi{http://dx.doi.org/}

\begin{document}
\title{Hydrodynamics of quantum entropies in Ising chains with linear 
dissipation}
\author{Vincenzo Alba}
\address{$^1$Institute  for  Theoretical  Physics, Universiteit van Amsterdam,
Science Park 904, Postbus 94485, 1098 XH Amsterdam,  The  Netherlands}
\author{Federico Carollo}
\address{$^2$Institut f\"ur Theoretische Physik, Universit\"at T\"ubingen,
	Auf der Morgenstelle 14, 72076 T\"ubingen, Germany}

\begin{abstract}
	We study the dynamics of quantum information and of quantum correlations  after a quantum quench, in 
	transverse field Ising chains subject to generic 
	linear dissipation. As we show, in the hydrodynamic limit of  long times, 
	large system sizes, and weak dissipation, 
	entropy-related quantities ---such as the von Neumann 
	entropy, the R\'enyi entropies, and the associated mutual information--- admit a 
	simple description within the so-called quasiparticle picture. Specifically, we analytically derive a hydrodynamic formula, recently conjectured for generic noninteracting 
	systems, which allows us to demonstrate a universal  feature of the dynamics of correlations in such dissipative noninteracting system. 
	For any possible dissipation,  
	the mutual information grows up to 
	a time scale that is proportional to the inverse dissipation rate, and then decreases, always vanishing in the long 
	time limit. In passing, we provide analytic formulas describing 
	the time-dependence of arbitrary functions of the fermionic covariance matrix, in the hydrodynamic limit. 
\end{abstract}

\maketitle

\section{Introduction}
\label{sec:intro}

Understanding the 
fate of entanglement and of quantum correlations in open 
quantum many-body systems is of paramount importance in order to assess the simulability of 
quantum devices with classical computers~\cite{oh2021classical,zhou2020what}, 
or to understand cold-atom experiments~\cite{islam2015measuring,kaufman2016quantum,brydges2019probing,elben2020mixed}.  
However, shedding light on these aspects of dissipative many-body quantum dynamics still represents a challenging task~\cite{rossini2021coherent}. 

For closed 
quantum many-body systems a powerful hydrodynamic picture, 
based on the existence of long-lived excitations, provides 
a thorough description of the spreading of entanglement, at least in 
integrable 
systems~\cite{calabrese2005evolution,fagotti2008evolution,alba2017entanglement,alba2018entanglement,alba2017quench,alba2017renyi,mestyan2018renyi,alba2019quantum}, 
both free (noninteracting) and interacting ones.  Unfortunately, much less is known in the presence of dissipation. Very recently, 
it has been shown that it is possible to extend the (hydrodynamic) quasiparticle picture in order to 
include the effects of dissipation in free-fermion and free-boson 
models~\cite{maity2020growth,alba2021spreading,carollo2021emergent} (see also Ref.~\cite{cao2019entanglement}). 
In particular, building on the results discussed in 
Ref.~\cite{alba2021spreading}, a formula, which allows to 
describe the dynamics of several entropy-related quantities, has been conjectured and thoroughly verified numerically
 \cite{carollo2021emergent} for rather general systems. 
 However, an analytic proof of this formula is still missing, even when considering  specific cases. The main contribution of this work is to analytically demonstrate the validity of the hydrodynamic picture, as described by the formula reported in 
Ref.~\cite{carollo2021emergent}, for the transverse field 
Ising chain subject to arbitrary linear dissipation.

\begin{figure}[t]
\begin{center}
\includegraphics[width=.9\textwidth]{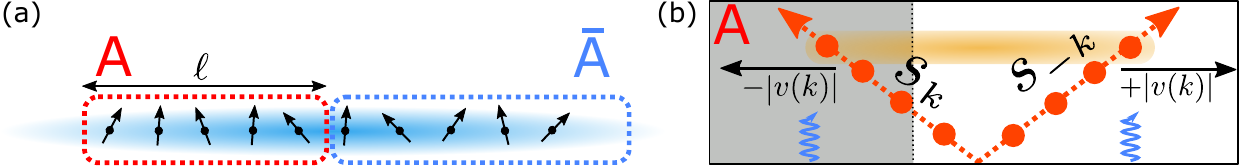}
\caption{(a) Setup considered in this work. We focus on the quantum 
 Ising chain subject to arbitrary linear dissipation acting on the full 
 system. We are interested 
 in the out-of-equilibrium dynamics of the subsystem entropies $S^{\scriptscriptstyle (n)}$ 
 and the mutual information ${\mathcal I}^{\scriptscriptstyle (n)}_{A:\bar A}$ 
 between a subsystem $A$ of length $\ell$ and its complement $\bar A$ after a generic 
 magnetic field quench. (b) 
 Quasiparticle (hydrodynamic) picture. 
 Pairs of correlated quasiparticles are produced after the quench. Quasiparticles 
 travel with opposite velocities $v(k)=v(-k)=-v(k)$, with $k$ the quasimomentum. Pairs that are 
 shared between $A$ and $\bar A$ contribute with $s_k$ and $s_{-k}$ to the  entropies. 
 Quantum entropies also contain 
 a contribution which is due to the system being in a mixed state [wiggly lines in (b)]. 
}
\end{center}
\label{fig0:cartoon}
\end{figure}

We consider open quantum many-body dynamics of Markovian type, for which the evolution of the full system density matrix $\rho$ is given by the 
Lindblad equation as~\cite{petruccione2002the}  
\begin{equation}
	\label{eq:liouv}
	\frac{d\rho}{dt}=-i[H,\rho]+\gamma\sum_{m}\Big(L_m\rho L_m^\dagger
	-\frac{1}{2}\{L_m^\dagger L_m,\rho\}\Big).
\end{equation}
Here, $H$ is the many-body Hamiltonian of the system, $\gamma$ represents an overall dissipation strength, and the operators $L_m$ encode 
how the presence of an environment affects the dynamics of the quantum system. We consider a bipartition of the many-body system into two complementary 
subsystems $A$ and $\bar A$, with $\bar A$ denoting the complement of $A$ [as 
illustrated in Fig.~\ref{fig0:cartoon}(a)], and study different entropy-related 
quantities. Namely, the R\'enyi entropies $S_A^{\scriptscriptstyle(n)}$ defined as~\cite{amico2008entanglement,calabrese2009entanglement,laflorencie2016quantum} 
\begin{equation}
	\label{eq:renyi-def}
	S_A^{(n)}:=\frac{1}{1-n}\ln(\mathrm{Tr}\rho_A^{n}), \quad n\in\mathbb{R},
\end{equation}
and the von Neumann entropy $S_A:=-\mathrm{Tr}\rho_A\ln(\rho_A)$. 
We further consider the mutual information ${\mathcal I}^{\scriptscriptstyle(n)}_{A:\bar A}$, 
which is given by 
\begin{equation}
	\label{eq:mi-def}
	{\mathcal I}^{(n)}_{A:\bar A}:=S_A^{(n)}+S_{\bar A}^{(n)}-S_{A\cup \bar A}^{(n)}. 
\end{equation}
For a system prepared in a pure state and undergoing unitary dynamics, it is easy to show that $S_A^{\scriptscriptstyle(n)}
=S_{\bar A}^{\scriptscriptstyle(n)}$ and $S_{A\cup\bar A}^{\scriptscriptstyle (n)}=0$. 
This is, however, not the case in the presence of dissipation since the total system 
$A\cup \bar A$ is, in general, in a mixed state. Furthermore, we recall here that for mixed states  the 
R\'enyi entropies and the mutual information are not proper measures of 
quantum entanglement. In particular, the mutual information only quantifies 
the total (classical plus quantum) correlation between $A$ and $\bar A$, which bounds the 
quantum entanglement between them~\cite{devetak2005distillation}. 
For mixed states, a proper measure of entanglement is given by the logarithmic negativity~\cite{vidal2002computable,plenio2005logarithmic,wichterich2009scaling,calabrese2012entanglement,eisler2005on,shapourian2019entanglement}.

Here we consider an arbitrary magnetic field quench 
in an open quantum Ising chain affected by a generic \emph{linear} dissipation. We rigorously show 
that the dynamics of the Re\'nyi entropy 
$S_A^{\scriptscriptstyle(n)}$ of an interval of length $\ell$ embedded in an 
infinite system is described by \cite{carollo2021emergent}
\begin{equation}
\label{eq:result}
S_A^{(n)}=\int_{-\pi}^{\pi}\frac{dk}{2\pi}\Big[s^{(n),YY}_k-s^{(n),\mathrm{mix}}_k
\Big]\min(\ell,2|v(k)|t)+\ell\int_{-\pi}^\pi\frac{dk}{2\pi}s^{(n),\mathrm{mix}}_k. 
\end{equation}
The above equation was recently conjectured in Ref.~\cite{carollo2021emergent}, 
and it has been verified numerically for several types of dissipation. 
Eq.~\eqref{eq:result} holds in what we call here the weakly-dissipative hydrodynamic limit characterized by  
$\ell,t\to\infty$, $\gamma\to0$ with $\ell/t$ and $\gamma\ell$ fixed,  
where $\gamma$ is the overall dissipation strength [see Eq.~\eqref{eq:liouv}] and $\ell$ is the size of the subsystem $A$ [cf.~Fig.~\ref{fig0:cartoon}]. 
The first term in~\eqref{eq:result} describes the 
spreading of correlations 
due to correlated pairs of quasiparticles traveling with velocities $v(k)$ and 
$v(-k)$. This term is similar to the one appearing in the 
unitary case, i.e., without dissipation. Remarkably, as we discuss, 
the group velocities $v(k)$ of the quasiparticles are not affected 
by the dissipation, and this could be attributed to the fact that the dissipation is
linear and weak. The term $s^{\scriptscriptstyle (n),YY}_k$ in the square 
brackets is reminiscent of the  Yang-Yang entropy in the unitary 
case~\cite{alba2021generalized}. The term $s_k^{\scriptscriptstyle(n),\mathrm{mix}}$ 
is purely dissipative. Indeed, for $\gamma=0$ and starting from a pure state, one has that $s_k^{\scriptscriptstyle 
(n),\mathrm{mix}}=0$, and $s_k^{\scriptscriptstyle(n),YY}$ becomes the 
Yang-Yang entropy. In contrast with the unitary case, in the presence of dissipation both $s_k^{\scriptscriptstyle (n),YY}$ 
and $s_k^{\scriptscriptstyle (n),\mathrm{mix}}$ are time-dependent. 
As it is clear from~\eqref{eq:result}, the effect of $s_k^{\scriptscriptstyle(n),\mathrm{mix}}$ 
is twofold. Specifically, it gives a volume-law contribution associated with the mixedness of the quantum state 
[see second term in~\eqref{eq:result}], and it reduces the total correlation 
between the quasiparticle pairs (term with the minus sign in~\eqref{eq:result}). 
Interestingly, in the limit $t\to\infty$ the term in the square 
brackets vanishes for any type of dissipation. We also provide a formula for 
the mutual information, which, in agreement with the result of Ref.~\cite{alba2021spreading} and 
\cite{carollo2021emergent}, depends only on the first term in~\eqref{eq:result}. 
Surprisingly, while for most of the dissipators the Yang-Yang 
term $s_k^{\scriptscriptstyle(n),YY}$ is determined by the dynamics of the 
density of quasiparticles that diagonalize the Ising chain, this is not true in 
general. We also show that, for those dissipations for which this is not true, $s_k^{\scriptscriptstyle(n),YY}$ 
is not an even function of $k$, in stark contrast with the usually considered Hamiltonian cases, and 
with the dissipators of Ref.~\cite{carollo2021emergent}. Finally, as a byproduct of our analysis we provide exact formulas 
describing the weakly-dissipative hydrodynamic limit of arbitrary functions of 
the fermionic covariance matrix of the Ising chain, generalizing the result of 
Ref.~\cite{calabrese2012quantum} to dissipative settings.

The manuscript is organized as follows. In section~\ref{sec:model} we 
introduce the quantum Ising chain (see section~\ref{sec:ising}), 
the quench protocol (section~\ref{sec:quench}), and the 
Lindblad framework that we exploit to account for dissipation~\ref{sec:lin}. In section~\ref{sec:K-par} 
we derive the behavior of the fermionic covariance matrix in the weakly-dissipative 
hydrodynamic limit. In section~\ref{sec:result} we present our main 
results, which are formulae~\eqref{eq:tr-F-man}~\eqref{eq:f-entropy}, 
and~\eqref{eq:mi-quasi}. Section~\ref{sec:quasi-rho} is devoted 
to the dynamics of the density of the quasiparticles, whereas 
in section~\eqref{sec:ss-vanishing} we discuss the steady-state value 
of the subsystem entropies. In section~\ref{sec:numerical} we 
provide numerical results. Specifically,  
we discuss numerical benchmarks for the entropy 
and the mutual information in section~\ref{sec:enta}, 
section~\ref{sec:mi-numerics}, and section~\ref{sec:unp}. 
Finally, we present our conclusions in section~\ref{sec:concl}. In~\ref{sec:obs} 
we review the calculation of subsystem entropies for free-fermion 
systems. In~\ref{sec:useful-1} and~\ref{sec:useful-2} we report a complete   
derivation of the results of section~\ref{sec:result}.

\section{Model, quench, and dissipation}
\label{sec:model}

In this paper, we are interested in the interplay between out-of-equilibrium unitary 
dynamics after a quantum quench~\cite{polkovnikov2011colloquium,calabrese2016introduction,vidmar2016generalized,essler2016quench} in the  
quantum Ising chain and the presence of arbitrary linear dissipation. 
In this section, we start by reviewing some relevant aspects 
of the Ising chain and 
then discuss the quantum quench protocol (magnetic field quench). Finally, in 
section~\ref{sec:lin} we introduce the Lindblad framework for Markovian open quantum dynamics. 

\subsection{Ising (Kitaev) chain}
\label{sec:ising}

We  consider the quadratic fermionic chain defined by the 
Hamiltonian 
\begin{equation}
	\label{eq:ham-is}
	H=-J\sum_{j=1}^L(c^\dagger_j c_{j+1}+c^\dagger_{j+1}c_j-\delta c_jc_{j+1}-
	\delta c^\dagger_{j+1}c^\dagger_j-2hc^\dagger_j c_j) -J h L. 
\end{equation}
Here $c_j$ are standard fermionic operators acting on site $j$ of the chain, 
and $L$ is the length of the chain. We use periodic boundary 
conditions $c_{j+L}=c_j$. 
In~\eqref{eq:ham-is} $h$ is the external magnetic field, and $J,\delta$ real parameters. 
In the following we are going to fix $\delta=1$, 
although our results could be extended to 
arbitrary $\delta$. We also fix $J=1$. The Hamiltonian~\eqref{eq:ham-is} is 
obtained from the transverse field Ising chain after a Jordan-Wigner transformation~\cite{sachdev2011quantum}. The total number of 
fermions $\hat N:=\sum_jc^\dagger_j c_j$ is not conserved for $\delta\ne0$, whereas 
the fermion parity $e^{i\pi \hat N}$ is a conserved quantitiy. Importantly, the 
Jordan-Wigner transformation is crucial 
in order to extract physical quantities both at equilibrium and out-of-equilibrium in 
the transverse field Ising chain. For instance, the Jordan-Wigner string 
affects the boundary conditions that have to be imposed 
on the fermions. Specifically, periodic boundary conditions on the spins are mapped 
to periodic or antiperiodic ones for the fermions, depending on the parity of the fermion 
number. 
Here, however, to avoid all these complications 
we work directly with the fermionic Hamiltonian~\eqref{eq:ham-is} with periodic boundary 
conditions.  

It is convenient to rewrite~\eqref{eq:ham-is} in terms of Majorana fermions. 
Let us define two species of Majorana operators $w_{1,j}$ and $w_{2,j}$ as 
\begin{equation}
	\label{eq:majo}
	w_{1,j}=c^\dagger_j+c_j,\quad w_{2,j}=i(c_j-c^\dagger_j), 
\end{equation}
with standard anticommutation relations 
\begin{equation}
	\{w_{s,j},w_{s',j'}\}=2\delta_{j,j'}\delta_{s,s'}. 
\end{equation}
It is convenient to think of these operators as the elements of a 
$2L$-dimensional operator-valued vector defined as $w_{2j-1}:=w_{1,j}$ and 
$w_{2j}:=w_{2,j}$. We will exploit this formalism when introducing the dissipative contribution to the dynamics below. 
By using~\eqref{eq:majo}, the Hamiltonian~\eqref{eq:ham-is} is recast as 
\begin{equation}
	\label{eq:ising-ham-2}
H=\sum_{j,m=1}^L(w_{1,j},\, w_{2,j})h_{j,m}\left(\begin{array}{c}
			w_{1,m}\\
			w_{2,m}
	\end{array}\right). 
\end{equation}
Here $h_{j,m}$ is the following $2\times 2$ matrix
\begin{equation}
	\label{eq:hjm-is}
	h_{j,m}=\frac{i}{2}\left(
		\begin{array}{cc}
			0 & \delta_{j,m+1}-h\delta_{j,m}\\
			-\delta_{j,m-1}+h\delta_{j,m} & 0
		\end{array}
	\right),
\end{equation}
specifying the (quadratic) interaction between site $j$ and site $m$. 
By exploiting translation invariance, Eq.~\eqref{eq:hjm-is} can be rewritten as 
\begin{equation}
h_{j,m}=\int_{-\pi}^{\pi}\frac{dk}{2\pi} e^{-i(j-m)k}\hat h_k, 
\label{blocks-h}
\end{equation}
where we introduce the so-called symbol $\hat h_k$ of the Hamiltonian 
\begin{equation}
	\label{eq:ham-symb}
\hat h_k=\frac{i}{2}\left(\begin{array}{cc}
			0 & e^{ik}-h\\
			-e^{-ik}+h &0
		\end{array}
	\right). 
\end{equation}
The fermionic Ising chain is diagonalized in terms of Bogoliubov modes $\alpha_k$. 
After a Bogoliubov transformation, indeed, Eq.~\eqref{eq:ising-ham-2} becomes 
\begin{equation}
	\label{eq:free}
	H=\sum_k\varepsilon_h(k)\alpha^\dagger_k\alpha_k. 
\end{equation}
Here $\alpha_k$ satisfy standard fermionic anticommutation relations, 
and $\varepsilon_h(k)$ is the single-particle dispersion relation
\begin{equation}
	\label{eq:vareps}
	\varepsilon_h(k)=2J\sqrt{1+h^2-2h\cos(k)}. 
\end{equation}
The Bogoliubov fermions $\alpha_k$ are defined in terms of the Fourier transforms 
of original fermions $c_k$ and $c^\dagger_{-k}$ as 
\begin{equation}
	\label{eq:rotation}
	(\alpha_k,\alpha_{-k}^\dagger)=\left(\begin{array}{cc}
			\cos(\frac{\theta_k}{2}) & i\sin(\frac{\theta_k}{2})\\
			i\sin(\frac{\theta_k}{2}) & \cos(\frac{\theta_k}{2})
	\end{array}\right)
	\left(\begin{array}{c}
			c_k\\
			c^\dagger_{-k}
	\end{array}\right)
\end{equation}
where $c_k:=L^{-1/2}\sum_{j=1}^L e^{i k j}c_j$. 
In~\eqref{eq:rotation}, we introduced the Bogoliubov angle $\theta_k$ defined through the relation 
\begin{equation}
	\label{eq:theta}
e^{i\theta_k}=\frac{h-e^{ik}}{\sqrt{1+h^2-2h\cos(k)}}. 
\end{equation}
A crucial ingredient for the following is the Majorana covariance matrix $\Gamma$. 
This is defined through the $2\times2$ blocks 
\begin{equation}
	\label{eq:gamma-def}
	\Gamma_{j,j'}=
	\left(\begin{array}{cc}
			\langle w_{1,j}w_{1,j'}\rangle-\delta_{j,j'} & \langle w_{1,j}w_{2,j'}\rangle\\
			\langle w_{2,j}w_{1,j'}\rangle & 	\langle w_{2,j}w_{2,j'}\rangle-\delta_{j,j'}	
	\end{array}\right)
	,\quad j,j'=1,\dots,L, 
\end{equation}
where $\langle x\rangle:=\mathrm{Tr}(\rho x)$, and $\rho$ is the density matrix that 
describes the state of the system. 
In translational invariant situations, $\Gamma$ depends only on $j-j'$ and it 
can be written in terms of its symbol $\hat \Gamma_k$ as 
\begin{equation}
	\label{eq:gamma-k-inv}
	\Gamma_{j,j'}(t)=\int_{-\pi}^{\pi}\frac{dk}{2\pi}e^{-ik(j-j')}\hat \Gamma_k(t). 
\end{equation}
It is well-known~\cite{peschel2009reduced} that from~\eqref{eq:gamma-def} it is 
possible to extract the von Neumann entropy and the R\'enyi entropies of 
a subsystem $A$ (see~\ref{sec:obs}). 

\subsection{Magnetic field quench}
\label{sec:quench}

The protocol of the magnetic field quench is as follows: 
At $t=0$ the chain is prepared in the ground state of~\eqref{eq:ham-is} 
at $h_0$. Then the magnetic field is suddenly changed to its final 
value $h_0\to h$. Since the initial state is not an eigenstate of the 
final Hamiltonian, in the absence of dissipation, the  chain undergoes 
nontrivial unitary dynamics. 
The generic magnetic field quench is parametrized in terms of a new 
Bogoliubov angle~\cite{calabrese2012quantum,calabrese2012quantumquench,caux2013time} 
$\Delta_k$. This is given as $\Delta_k:=\theta_{k}-\theta_k^0$, 
where $\theta_k$ is defined in~\eqref{eq:theta} and $\theta_k^0$ is 
obtained from~\eqref{eq:theta} by setting $h\to h_0$. The angle $\Delta_k$ 
enters in the transformation between the Bogoliubov modes $\alpha_k$ and 
$\alpha_k^0$ diagonalizing the Ising Hamiltonian with magnetic field $h$ and 
$h_0$, respectively. 
Both modes $\alpha_k$ and $\alpha_k^0$ are written in terms of 
the same fermions $c_k$, implying that the angle $\Delta_k$ is easily obtained 
from~\eqref{eq:rotation}. Specifically, one obtains that 
\begin{align}
	\label{eq:cD}
	& \cos(\Delta_k)=4\frac{1+h_0 h+\cos(k)(h+h_0)}{\varepsilon_{h_0}(k)\varepsilon_h(k)}\\
	\label{eq:sD}
	& \sin(\Delta_k)=4\frac{\sin(k)(h_0-h)}{\varepsilon_h(k)\varepsilon_{h_0}(k)}. 
\end{align}
Here $\varepsilon_h$ is the dispersion of the Bogoliubov quasiparticles (cf.~\eqref{eq:vareps}). 

We are interested in the dynamics of the covariance matrix $\Gamma$ (cf.~\eqref{eq:gamma-def}). 
First, the ground-state pre-quench matrix $\Gamma(0)$ can be derived explicitly~\cite{calabrese2012quantumquench}. The associated symbol $\hat \Gamma_k$ reads as 
\begin{equation}
	\label{eq:g0-sy}
	\hat \Gamma_k(0)=-i\left(
		\begin{array}{cc}
			0 & -e^{i\theta_k+i\Delta_k}\\
			e^{-i\theta_k-i\Delta_k} & 0
		\end{array}
\right),
\end{equation}
where $\Delta_k$ and $\theta_k$ are defined in~\eqref{eq:cD}~\eqref{eq:sD} 
and~\eqref{eq:theta}. 
The evolved symbol $\Gamma_k(t)$ is obtained from~\eqref{eq:g0-sy} as 
\begin{equation}
	\label{eq:gamma-ev}
	\hat \Gamma_k(t)=\hat U_k \hat \Gamma_k(0)\hat U_{-k}^T. 
\end{equation}
Here $T$ denotes matrix transposition, and the symbol of the evolution 
operator $\hat U_k$ is given  as 
\begin{equation}
	\label{eq:evol}
	\hat U_k=e^{-4i \hat h_k t}, 
\end{equation}
with $\hat h_k$ the symbol of the Hamiltonian~\eqref{eq:ham-symb}. 
More explicitly, by using~\eqref{eq:g0-sy}\eqref{eq:gamma-ev}\eqref{eq:evol} 
we obtain the compact expression for $\hat\Gamma_k(t)$ as 
\begin{equation}
	\label{eq:symb-no-diss}
	\hat\Gamma_k(t)=-\cos(\Delta_k)\sigma_y^{(k)}-\sin(\Delta_k)\sigma_x^{(k)}e^{2i\varepsilon_h(k)t\sigma_{y}^{(k)}}. 
\end{equation}
At $t=0$ one recovers~\eqref{eq:g0-sy}. Note that only the second 
term in~\eqref{eq:symb-no-diss} depends on time and that
in~\eqref{eq:symb-no-diss} we have introduced the rotated 
Pauli matrices $\sigma^{(k)}_\alpha$ as 
\begin{equation}
\label{eq:Pauli-rotated}
\sigma^{(k)}_\alpha:= e^{i\theta_k\sigma_z/2}\sigma_\alpha e^{-i\theta_k\sigma_z/2},\quad \alpha=x,y,z,
\end{equation}
where $\sigma_\alpha$ with $\alpha=x,y,z$ are the standard Pauli matrices. 
The definition~\eqref{eq:Pauli-rotated} will be convenient in section~\ref{sec:K-par} 
to derive the dynamics of $\hat \Gamma_k(t)$ in the presence of dissipation.  
To obtain~\eqref{eq:symb-no-diss} we used the commutation relations of 
the Pauli matrices and that 
$[\sigma_y^{\scriptscriptstyle(-k)}]^T=-\sigma_y^{\scriptscriptstyle(k)}$. 

The Majorana covariance matrix $\Gamma$ is obtained as the inverse Fourier transform 
of~\eqref{eq:symb-no-diss} as 
\begin{equation}
\Gamma=-i\left(
		\begin{array}{cc}
			f_{j-j'}(t) & -g_{-(j-j')}(t)\\
			g_{j-j'}(t) & -f_{j-j'}(t)
	\end{array}\right),
\end{equation}
where we defined the functions $f_{j-j'}$ and $g_{j-j'}$ as 
\begin{align}
	f_l(t)&=i\int_{-\pi}^\pi \frac{dk}{2\pi}e^{-ikl}
	\sin(\Delta_k)\sin(2\varepsilon_h(k)t),\\	
	g_l(t)&=-\int_{-\pi}^\pi\frac{dk}{2\pi}e^{-ikl}e^{-i\theta_k}
	\Big[\cos(\Delta_k)-i\sin\Delta_k \cos(2\varepsilon_h(k)t)\Big]. 
\end{align}

In this work we focus on the thermodynamic limit and on the long-time 
limit after the quench. In this limit, the reduced-density matrix of 
any finite subystem $A$ (see Fig.~\ref{fig0:cartoon}) reaches a 
steady state that is described by a Generalized Gibbs Ensemble 
(GGE)~\cite{calabrese2016introduction,essler2016quench,vidmar2016generalized}. 
The GGE is identified by the occupations (densities) $\rho_k$ of the modes $\alpha_k$ 
(cf.~\eqref{eq:free}). These are obtained as the expectation value 
\begin{equation}
	\rho_k:=\langle h_0|\alpha^\dagger_k\alpha_k|h_0\rangle, 
\end{equation}
where $|h_0\rangle$ is the pre-quench initial state. The density $\rho_k$ is 
preserved during the unitary dynamics after  the quench. Crucially, this is 
not the case in the presence of dissipation. We will derive the dynamics of 
$\rho_k$ in the presence of dissipation in section~\ref{sec:K-par}. 
By using the explicit form of the operators $\alpha_k$ one obtains 
that~\cite{calabrese2012quantum} 
\begin{equation}
	\label{eq:rhok}
	\rho_k=\frac{1-\cos(\Delta_k)}{2}. 
\end{equation}
It is useful to stress  that 
by using~\eqref{eq:rotation} it is possible to express the Bogoliubov modes 
$\alpha_k$ in terms of Majorana operators $w_{s,j}$ (cf.~\eqref{eq:majo}), 
which allows to rewrite $\rho_k$ in terms of 
the matrix elements of $\hat \Gamma_k$. The result reads as~\cite{caux2013time}  
\begin{equation}
	\label{eq:rho-gamma}
	\rho_k=\frac{1}{2}(1-\mathrm{Im}(e^{-i\theta_k}[\hat\Gamma_k]_{12})),
\end{equation}
where $[\hat\Gamma_k]_{12}$ is the only independent off-diagonal entry
of $\hat \Gamma_k(t)$ (cf.~\eqref{eq:gamma-ev}). 
In the presence of dissipation Eq.~\eqref{eq:rho-gamma} provides a way 
to extract how the density of modes $\alpha_k$ evolves under the combined 
action of unitary and dissipative terms. 

\subsection{Lindblad evolution of the covariance matrix}
\label{sec:lin}

Here we consider a generic Liovillian dynamics (see~\eqref{eq:liouv}) 
for the Ising chain~\eqref{eq:ham-is} subject to 
arbitrary one-site translation invariant linear Lindblad operators. 
This type of dissipation can be treated within the framework of the so-called 
third quantization~\cite{prosen2008third}. 
Instead of considering the dynamics of the system density matrix~\eqref{eq:liouv}, 
it is convenient to focus on a generic observables $X_t$.  
Eq.~\eqref{eq:liouv} can be recast in the convenient form 
\begin{equation}
	\label{eq:liouv-1}
	\frac{d X_t}{dt}=i[H,X_t]+\sum_{m,j=1}^{2L} K_{mj}
	\left(w_m X_t w_j-\frac{1}{2}\left\{X_t,w_mw_j\right\}\right), 
\end{equation}
where $w_j$ are Majorana operators, $H$ is the system Hamiltonian written in 
the Majorana basis. Eq.~\eqref{eq:liouv-1} is written in terms of the 
$2L$ Majorana fermions $w_{2j-1}:=w_{1,j}$ and $w_{2j}:=w_{2,j}$ (cf.~\eqref{eq:majo}). 
In~\eqref{eq:liouv-1} the dissipation is encoded in the so-called 
Kossakowski matrix~\cite{gorini1976completely,lindblad1976on} 
$K_{mj}$. We make here a remark about the notation of the paper. With the writing $K_{mj}$, as in the above equation, we indicate the element of the $2L\times 2L$ matrix $K$ in the $m$th row and $j$th column. With the writing $K_{m,j}$, for instance appearing in Eq.~\eqref{blocks-h} for $h$, instead, we denote the $2\times2$ block of $K$ in the $m,j$ position.

Importantly, since the evolution implemented by Eq.~\eqref{eq:liouv-1}  has to be completely positive (see~\cite{petruccione2002the}), 
the Kossakowski matrix $K_{mj}$ has to be positive semi-definite~\cite{lindblad1976on}. 
For later convenience, we can decompose $K$ as 
\begin{align}
	\label{eq:re}
	& K^{\mathrm{Re}}:=\frac{1}{2}(K+K^*)=(K^{\mathrm{Re}})^T\\
	\label{eq:im}
	& K^{\mathrm{Im}}:=\frac{1}{2i}(K-K^*)=-(K^{\mathrm{Im}})^T, 
\end{align}
where we suppressed the indices in $K_{mj}$ to lighten the notation. The superscripts 
$\mathrm{Re}$ and $\mathrm{Im}$ denote the real and imaginary parts of $K$, respectively. 
Again, for translation invariant dissipation, by going to momentum space, one 
defines the symbol $\hat K_k$ of $K$ through the relation
\begin{equation}
	\label{eq:Ko-k}
	K_{m,j}=\int_{-\pi}^{\pi}\frac{dk}{2\pi}e^{-ik(m-j)}\hat K_k,
\end{equation}
where we remark one more time that $K_{m,j}$ is a $2\times2$ block of $K$. 
We note that for the positivity of $K$, one has that  
 $\hat K_k$ has to be positive. 
Using Eq.~\eqref{eq:liouv-1}  on products of Majorana operators in order to find the dynamics 
of $\Gamma$, one gets the differential equation~\cite{carollo2021emergent}  
\begin{equation}
	\label{eq:gamma-evol}
	\frac{d}{dt}\Gamma=\Lambda \Gamma +\Gamma \Lambda^{T}+4iK^{\mathrm{Im}}. 
\end{equation}
Here we defined the matrix $\Lambda$ to be 
\begin{equation}
	\label{eq:lambda}
	\Lambda:=-4i h-2K^{\mathrm{Re}},
\end{equation}
with $h$ the Hamiltonian matrix in the basis of Majorana defined in~\eqref{eq:hjm-is}. 
Interestingly, as it is clear from~\eqref{eq:gamma-evol}, only the real part 
$K^{\mathrm{Re}}$ enters in the matrix $\Lambda$, whereas $K^{\mathrm {Im}}$ acts as 
a driving term. Eq.~\eqref{eq:gamma-evol} has the formal solution 
\begin{equation}
\label{eq:gamma-evol-sol}
\Gamma(t)=\widetilde U(t)\Gamma(0)\widetilde U^T(t)+
4i\int_0^t du\, \widetilde U(t-u) K^{\mathrm{Im}}\widetilde U^T(t-u)\, ,
\end{equation}
where the evolution operator $\widetilde U(t)$ is defined as $\widetilde U(t):=e^{\Lambda t}$. 
For translation invariant systems, Eq.~\eqref{eq:gamma-evol-sol} can be 
recast as an equation for the symbol $\hat \Gamma_k$ (cf.~\eqref{eq:gamma-k-inv}). 
Specifically, one obtains that 
\begin{equation}
\label{eq:gamma-symb}
\hat \Gamma_k(t)={\widetilde U}_k\hat \Gamma_k(0)\widetilde U^T_{-k}(t) +
4i\int_{0}^t du \widetilde U_k(t-u)\hat K_k^{\mathrm{Im}} \widetilde U^T_{-k}(t-u). 
\end{equation}
Here $\hat \Gamma_k,\hat \Lambda_k,\hat K_k^{\mathrm{Im}}$ are the symbols of the 
matrices $\Gamma,\Lambda,K^{\mathrm{Im}}$, and are $2\times 2$ matrices. 
The integral in Eq.~\eqref{eq:gamma-symb} can be performed explicitly, although for generic 
dissipation the result is quite cumbersome. Still, Eq.~\eqref{eq:gamma-symb} is 
useful to obtain exact numerical results for $\hat\Gamma_k$. 
Further simplifications occur in the hydrodynamic limit $t\to\infty$ with 
weak dissipation, as we are going to show.

To proceed, let us consider the most general Kossakowski matrix $K$. 
Since its symbol $\hat K_k$ has to be hermitian (actually positive), it has to be of the  
form 
\begin{equation}
	\label{eq:K-symb}
	\hat K_k=\gamma\left(\begin{array}{cc}
			a(k) & c(k)\\
			c^*(k) & b(k)
		\end{array}
	\right).
\end{equation}
In~\eqref{eq:K-symb} we re-introduced the strength of the dissipation $\gamma$. 
The weak-dissipation limit 
corresponds to $\gamma\to0$ with an appropriate scaling with $\ell$. In~\eqref{eq:K-symb} $a(k),b(k)$ are real functions, 
whereas $c(k)$ is in general complex. 
From~\eqref{eq:K-symb} one can obtain the symbols 
$\hat K_k^{\mathrm{Re}}$ and $\hat K_k^{\mathrm{Im}}$ of 
the real and imaginary parts of $K$ as 
\begin{equation}
	\label{eq:k-re-im}
	\hat K_k^{\mathrm{Re}}=\gamma\left(
		\begin{array}{cc}
			a_{\mathrm{e}} & c^{\mathrm{Re}}_{\mathrm{e}}+i c^{\mathrm{Im}}_{\mathrm{o}}\\
			c^{\mathrm{Re}}_{\mathrm{e}}-i c^{\mathrm{Im}}_{\mathrm{o}} & b_{\mathrm{e}}
		\end{array}
	\right),\quad
	\hat K_k^{\mathrm{Im}}=\gamma\left(
		\begin{array}{cc}
			-i a_{\mathrm{o}} & -i c^{\mathrm{Re}}_{\mathrm{o}}+c^{\mathrm{Im}}_{\mathrm{e}}\\
			-i c^{\mathrm{Re}}_{\mathrm{o}}-c^{\mathrm{Im}}_{\mathrm{e}} & -i b_{\mathrm{o}}
		\end{array}
	\right), 
\end{equation}
where $c(k)=c^{\mathrm{Re}}(k)+ic^{\mathrm{Im}}(k)$, and 
we introduced the odd and even functions 
$a_{\mathrm{e}}(k):=[a(k)+a(-k)]/2$, $a_{\mathrm{o}}:=[a(k)-a(-k)]/2$. 
Similar definitions hold  for $b_\mathrm{e/o}(k)$ and 
$c^{\mathrm{Re/Im}}_{\mathrm{e/o}}(k)$. 
The decomposition~\eqref{eq:k-re-im} is obtained from~\eqref{eq:Ko-k} by 
using that $K^{\mathrm{Re}}=(K+K^*)/2$ and $K^{\mathrm{Im}}=(K-K^*)/(2i)$. 
We note that the fact that the symbol~\eqref{eq:K-symb} has to be positive semi-definite 
for any $k$ puts some constraints on the functions $a,b,c$. 
Interestingly, however, the results that we derive below hold true even when $\hat{K}$ is not positive, i.e. for ``unphysical" dissipation.

For the sake of concreteness, it is useful to specialize~\eqref{eq:k-re-im} to the case of gain/loss 
dissipation. This has been studied extensively in the literature, as it is 
one of the simplest yet experimentally relevant sources of 
dissipation~\cite{rossini2021coherent} (see~\cite{nigro2019competing} 
for a study of the interplay between criticality and dissipation, 
and~\cite{bouchoule2020the} for an application to cold-atom systems). 
The dynamics of the von Neumann entropy and of the mutual information 
in the presence of gain/loss dissipation 
in a simple tight-binding chain has been obtained analytically using the 
quasiparticle picture in Ref.~\cite{alba2021spreading} 
(see also Ref.~\cite{maity2020growth} for a study in the 
Ising chain). The case of  
localized losses has been addressed in Ref.~\cite{alba2021unbounded}. 
We consider gain/loss dissipation  with rates $\gamma g^\pm$, where $\gamma$ is the strength 
of the dissipation, and $g^\pm$ real parameters. The Lindblad operators $L_m^\pm$ 
are given as $L_m^{+}=2\sqrt{\gamma g^+}c^\dagger_m$ and 
$L_m^-=2\sqrt{\gamma g^-}c_m$. These operator model the incoherent creation and loss 
of fermions in the chain. 
By using~\eqref{eq:Ko-k}, one obtains that the 
symbol of the Kossakowski matrix $\hat K_k$ for gain/loss dissipation reads 
as 
\begin{equation}
	\label{eq:K-gl}
	\hat K_k=\gamma\left(\begin{array}{cc}
			g^++g^- & - i(g^--g^+)\\
			+i(g^--g^+)& g^++g^-
	\end{array}\right). 
\end{equation}
Note that $\hat K_k$ does not depend on the 
momentum $k$. This is a feature of dissipations which are diagonal in the lattice space. 
As it is clear from~\eqref{eq:K-gl}, 
$\hat K^{\mathrm{Re}}_k=(g^++g^-)\mathds{1}_2$,  
where $\mathds{1}_2$ is the $2\times 2$ identity matrix. This implies that the dynamics is 
determined by the model without dissipation (cf.~\eqref{eq:lambda}), apart from an 
exponential damping factor $e^{-2\gamma(g^++g^-)t}$. The driving term  
(cf.~\eqref{eq:gamma-symb}) $\hat K^{\mathrm{Im}}_k$ is purely imaginary, 
$k$-independent, and antisymmetric. 

\section{Covariance matrix in the weakly-dissipative hydrodynamic limit}
\label{sec:K-par}

In this section we derive a compact formula for the time-dependent symbol $\hat \Gamma_k$ of the 
Majorana correlator after a magnetic field quench (see section~\ref{sec:quench}) 
in the Ising chain in the presence of arbitrary 
dissipation. We focus on the dissipative hydrodynamic limit 
with $\gamma\to0$ (cf.~\eqref{eq:K-symb}), $t\to\infty$ with 
$\gamma t$ fixed. 

First,  in constructing the evolution operator 
$e^{\Lambda_k t}$, it is useful to introduce the modified dispersion 
relation $\tilde\varepsilon(k)$ as 
\begin{equation}
	\label{eq:teps}
	\tilde \varepsilon(k)=2\sqrt{1+h^2-2h\cos(k)+
	2i\gamma[(h-\cos(k))c^{\mathrm{Im}}_{\mathrm{o}}+c^{\mathrm{Re}}_{\mathrm{e}}\sin(k)]},
\end{equation}
where $c_\mathrm{o}^{\mathrm{Im}}$ and $c_\mathrm{e}^{\mathrm{Re}}$ are defined 
in~\eqref{eq:k-re-im}. In~\eqref{eq:teps} we are neglecting terms 
${\mathcal O}(\gamma^2)$ because they are 
irrelevant in the weak-dissipation  limit. Note that due to the term proportional 
to $\gamma$, $\tilde\varepsilon(k)$ is not an even function of $k$, in contrast with the 
unitary case. It is useful to expand~\eqref{eq:teps} in the limit $\gamma\to0$ to obtain 
\begin{equation}
	\label{eq:expansion}
	i\tilde\varepsilon(k)-i\tilde\varepsilon(-k)=	
	4\gamma z_{\mathrm{o}}(k)
	+{\mathcal O}(\gamma^2). 
\end{equation}
where the term ${\mathcal O}(\gamma^2)$ is irrelevant in the scaling limit. 
In~\eqref{eq:expansion} we have introduced the function $z_\mathrm{o}(k)$, which depends on the 
 the Bogoliubov angle $\theta_k$ defined in~\eqref{eq:theta}, as 
\begin{equation}
	\label{eq:z-def}
z_\mathrm{o}(k):=\cos(\theta_k)c_{\mathrm{o}}^{\mathrm{Im}}
	+\sin(\theta_k)c_{\mathrm{e}}^{\mathrm{Re}}. 
\end{equation}
As it is clear from its definition, $z_\mathrm{o}(k)$ is an odd function 
of $k$ and is in general  nonzero. We will show in section~\ref{sec:result} 
that this has striking consequences for the dynamics of the subsystem  entropies. 

To proceed, we observe that the symbol of the evolution operator 
$\widetilde U_k$ (cf.~\eqref{eq:gamma-symb}) 
in the weakly-dissipative scaling limit  takes the quite simple form 
\begin{equation}
\label{eq:u-diss}
\widetilde U_k=e^{-\gamma t(a_{\mathrm{e}}+b_{\mathrm{e}})}\exp\big(-i\tilde\varepsilon(k)\sigma_y^{(k)}t\big),
\quad
\widetilde U^T_{-k}=e^{-\gamma t(a_{\mathrm{e}}+b_{\mathrm{e}})}\exp\big(i\tilde\varepsilon(-k)\sigma_y^{(k)}t\big). 
\end{equation}
In~\eqref{eq:u-diss} the matrices $\sigma_\alpha^{(k)}$ are the rotated Pauli matrices 
introduced in~\eqref{eq:Pauli-rotated}. 
We are now ready to derive the time-dependent symbol~\eqref{eq:gamma-symb}. 
We decompose $\hat\Gamma_k(t)$ as 
\begin{equation}
	\hat\Gamma_k(t)=\hat\Gamma_k^{(1)}(t)+\hat\Gamma_k^{(2)}(t),
\end{equation}
where $\hat\Gamma_k^{(1)}$ and $\hat\Gamma^{(2)}_k$ correspond to the 
first and second term in~\eqref{eq:gamma-symb}, respectively. 
Let us first focus on $\Gamma_k^{\scriptscriptstyle(1)}$. 
The symbol of the initial correlator $\hat\Gamma_k(0)$ 
(cf.~\eqref{eq:g0-sy}) can be written as 
\begin{equation}
	\label{eq:g0-dec}
	\hat\Gamma_k(0)=-\cos(\Delta_k)\sigma_y^{(k)}-\sin(\Delta_k)\sigma_x^{(k)}, 
\end{equation}
where $\Delta_k$ is the Bogoliubov angle defined in~\eqref{eq:cD}\eqref{eq:sD}. 
$\hat\Gamma^{(1)}$ is obtained by applying~\eqref{eq:u-diss} in~\eqref{eq:gamma-symb}. 
This yields 
\begin{equation}
	\label{eq:gamma-1}
	\hat\Gamma_k^{(1)}=-e^{-2\gamma(a^e+b^e)t}(\cos(\Delta_k) \sigma_y^{(k)} 
		e^{i(\tilde\varepsilon(-k)-\tilde\varepsilon(k))t\sigma_y^{(k)}}+\sin(\Delta_k)
	\sigma_x^{(k)} e^{i(\tilde\varepsilon(-k)+\tilde\varepsilon(k))t\sigma_y^{(k)}}), 
\end{equation}
where we used the identity
\begin{equation}
	e^{-i\tilde\varepsilon(k)t\sigma_y^{(k)}}\sigma_{\alpha}^{(k)}
	e^{i\tilde\varepsilon(-k)t\sigma_y^{(k)}}=\sigma_{\alpha}^{(k)}
	e^{i(\tilde\varepsilon(-k)+\tilde\varepsilon(k))t\sigma_y^{(k)}},\quad\alpha=x,z. 
\end{equation}
In the absence of dissipation $\gamma\to0$, one recovers~\eqref{eq:symb-no-diss}. 
Interestingly, the condition $z_{\mathrm o}(k)=0$ (cf.~\eqref{eq:z-def}) implies that 
$\tilde\varepsilon(k)-\tilde\varepsilon(-k)=0$. Together with the fact 
that $\tilde\varepsilon(k)+\tilde\varepsilon(-k)=2\varepsilon_h(k)$, this 
implies that $\Gamma_k^{\scriptscriptstyle(1)}$ coincides with the 
time-evolved correlator~\eqref{eq:symb-no-diss} in the case without dissipation, 
apart from the overall damping factor.  
On the other hand, for nonzero $z_{\mathrm o}$, 
Eq.~\eqref{eq:gamma-1} contains a term proportional to 
$\sigma_y^{(k)}$, i.e., as in~\eqref{eq:symb-no-diss}, 
and a time-dependent term proportional to the identity $\mathds{1}_2$. 
This is absent in the unitary case, and it plays an important role in the dynamics 
(see section~\ref{sec:result}). We can recast~\eqref{eq:gamma-1} as  
\begin{multline}
	\label{eq:gamma-1b}
	\hat\Gamma_k^{(1)}=-e^{-2\gamma(a^e+b^e)t}(\cosh(\Delta_k) 
		\sinh(4z_\mathrm{o}\gamma t)\mathds{1}_2
		\\+\cosh(\Delta_k)\cosh(4z_\mathrm{o}\gamma t)\sigma_y^{(k)}
	+\sin(\Delta_k)\sigma_x^{(k)} e^{2i\varepsilon_h(k)t\sigma_y^{(k)}}), 
\end{multline}
where we used that $\tilde\varepsilon(k)+\tilde\varepsilon(-k)=2\varepsilon_h(k)$, with 
$\varepsilon_h(k)$ the dispersion of the Ising chain without dissipation~\eqref{eq:vareps}, 
and $z_\mathrm{o}(k)$ defined in~\eqref{eq:z-def}. Interestingly, 
all the terms in~\eqref{eq:gamma-1b} 
depend on time, although in the first two the time dependence is 
only through the scaling variable $\gamma t$. This means that in the 
hydrodynamic limit they can be treated as constants. 
This is not the case for the last term in~\eqref{eq:gamma-1b}, which governs the 
dynamics of the correlated quasiparticles created after the quench. 
Interestingly, the fact that the dispersion $\varepsilon_h(k)$ of the 
Ising chain appears in~\eqref{eq:gamma-1} implies that the 
velocity $v(k)=d\varepsilon_h(k)/dk$ of the quasiparticles is not affected by 
the dissipation, at least in this hydrodynamic limit. 

Let us now consider the term $\hat\Gamma_k^{\scriptscriptstyle(2)}$, i.e., the second 
term in~\eqref{eq:gamma-symb}. One first rewrites $\hat K_k^{\mathrm{Im}}$ as 
\begin{equation}
	\hat K_k^{\mathrm{Im}}=
	-i\gamma\Big[\frac{a_{\mathrm{o}}+b_{\mathrm{o}}}{2}\mathds{1}_2+
	\frac{a_{\mathrm{o}}-b_{\mathrm{o}}}{2}
	\sigma_z+c_{\mathrm{o}}^{\mathrm{Re}}\sigma_x
-c_{\mathrm{e}}^{\mathrm{Im}}\sigma_y\Big], 
\end{equation}
where the functions $a_{\alpha},b_{\alpha},
c_{\alpha}^{\mathrm{Re}},c_{\alpha}^{\mathrm{Im}}$ ($\alpha=\mathrm{e},\mathrm{o}$) 
are defined in~\eqref{eq:k-re-im}. It is useful to rewrite $\hat K_k^{\mathrm{Im}}$ 
in terms of the rotated Pauli matrices $\sigma_\alpha^{\scriptscriptstyle(k)}$ 
(cf.~\eqref{eq:Pauli-rotated}) as 
\begin{multline}
	\label{eq:Kk}
	\hat K_k^{\mathrm{Im}}=-i\gamma\Big[
	\frac{a_{\mathrm{o}}+b_{\mathrm{o}}}{2}\mathds{1}_2+
	\frac{a_{\mathrm{o}}-b_{\mathrm{o}}}{2}
	\sigma_z^{(k)}+\\
	(c^{\mathrm{Re}}_{\mathrm{o}}\cos(\theta_k)+
	c^{\mathrm{Im}}_{\mathrm{e}}\sin(\theta_k))
	\sigma^{(k)}_x+(c^{\mathrm{Re}}_{\mathrm{o}}\sin(\theta_k)-
c^{\mathrm{Im}}_{\mathrm{e}}\cos(\theta_k))\sigma^{(k)}_y\Big]. 
\end{multline}
After substituting~\eqref{eq:Kk} in~\eqref{eq:gamma-symb}, it is straightforward to 
verify that the second and third terms in~\eqref{eq:Kk} give contributions ${\mathcal O}(\gamma)$, 
which are vanishing in the hydrodynamic limit. Only the 
first and last term in~\eqref{eq:Kk} give 
a finite contribution. To proceed, we use the  identities 
\begin{align}
	\label{eq:int-1}
	&\int_0^t du e^{-4\gamma(t-u)s} e^{4z\gamma(t-u)\sigma_y^{(k)}}= 
	\frac{1}{4\gamma}\frac{s\mathds{1}_2+
	z\sigma_y^{(k)}}{s^2-z^2}\Big[\mathds{1}_2-e^{-4\gamma st}e^{4\gamma z t\sigma_y^{(k)}}\Big]\\
	\label{eq:int-2}
	&\int_0^t du e^{-4\gamma(t-u)s} \sigma_y^{(k)}e^{4z\gamma(t-u)\sigma_y^{(k)}}= 
	\frac{1}{4\gamma}\frac{z\mathds{1}_2+
	s\sigma_y^{(k)}}{s^2-z^2}\Big[\mathds{1}_2-e^{-4\gamma st}e^{4\gamma z t\sigma_y^{(k)}}\Big], 
\end{align}
where $s,z$ are arbitrary numbers. 
Eq.~\eqref{eq:int-1} and~\eqref{eq:int-2} differ from 
each other by an exchange $s\leftrightarrow z$ in the numerator of the 
multiplicative factor. Importantly, Eq.~\eqref{eq:int-1} and~\eqref{eq:int-2} 
are proportional to $1/\gamma$. This means that they give a finite contribution in  
the weakly-dissipative hydrodynamic limit because the $1/\gamma$ cancels out 
the factor $\gamma$ in~\eqref{eq:Kk}. 
By applying~\eqref{eq:int-1} and~\eqref{eq:int-2} in~\eqref{eq:Kk}, we obtain that 
\begin{multline}
	\label{eq:gamma-2}
	\Gamma^{(2)}_k=\frac{e^{-4\gamma s_e t}}{s_\mathrm{e}^2-z_\mathrm{o}^2}\Big[
		(e^{4\gamma s_et}-\cosh(4\gamma z_\mathrm{o}t)(s_\mathrm{e} s_\mathrm{o}
			+z_\mathrm{e} z_\mathrm{o})-(z_\mathrm{e}s_\mathrm{e}+s_\mathrm{o}
			z_\mathrm{o})
		\sinh(4\gamma z_\mathrm{o}t))\mathds{1}_2\\+
		(e^{4\gamma s_\mathrm{e}t}-\cosh(4\gamma z_\mathrm{o}t)
			(s_\mathrm{e} z_\mathrm{e}+s_\mathrm{o} z_\mathrm{o})-
			(z_\mathrm{e}z_\mathrm{o}+s_\mathrm{e} s_\mathrm{o})
		\sinh(4\gamma z_\mathrm{o}t))\sigma_y^{(k)} 
	\Big]. 
\end{multline}
Here we redefined 
\begin{align}
	\label{eq:se-def}
	& s_\mathrm{e}:=\frac{a_{\mathrm{e}}+b_{\mathrm{e}}}{2}\\
	\label{eq:so-def}
	& s_\mathrm{o}:=\frac{a_{\mathrm{o}}+b_{\mathrm{o}}}{2}\\
	\label{eq:ze-def}
	& z_\mathrm{e}:=c^{\mathrm{Re}}_{\mathrm{o}}\sin(\theta_k)-
	c^{\mathrm{Im}}_{\mathrm{e}}\cos(\theta_k),
\end{align}
and $z_\mathrm{o}$ is defined in~\eqref{eq:z-def}. 
Putting together~\eqref{eq:gamma-1b} and~\eqref{eq:gamma-2} one obtains the symbol 
$\hat\Gamma_k$ for the most general linear translation-invariant 
dissipation as 
\begin{equation}
	\label{eq:symbolo}
	\hat\Gamma_k=C_k\mathds{1}_2+A_k\sigma_y^{(k)}+B_k\sigma_x^{(k)}
	e^{2i\varepsilon_h(k)t\sigma_y^{(k)}}, 
\end{equation}
where the functions $A_k,B_k,C_k$ are obtained from~\eqref{eq:gamma-1b} and~\eqref{eq:gamma-2}. 
Clearly, $C_k,B_k$ are odd functions of $k$, whereas $A_k$ is an even one. This is in accord 
with $\mathrm{Tr}(\Gamma)=\int dk\mathrm{Tr}(\hat\Gamma_k)=0$, which holds by definition for 
$\Gamma$.

\section{Quantum entropies in the weakly-dissipative hydrodynamic limit} 
\label{sec:result}

We now discuss our main result showing that it is possible to describe 
analytically the dynamics of the von Neumann and the R\'enyi entropies, in the 
discussed hydrodynamic limit, for any given subsystem 
$A$ of length $\ell$ (see Fig.~\ref{fig0:cartoon}). 
Again, the limit is defined as $t,\ell\to\infty$, 
$\gamma\to0$ and $t/\ell$ and $\gamma t$ fixed. 

Let us start by introducing the $2\ell\times 2\ell$ matrix $\Gamma_\ell$ of the 
Majorana correlations restricted to $A$. As we already discussed in 
section~\ref{sec:K-par}, the symbol $\hat \Gamma_k$ is of the 
form~\eqref{eq:symbolo}. The subscript $\ell$ in $\Gamma_\ell$ is to stress that 
the correlator is restricted to subystem $A$ (see Fig.~\ref{fig0:cartoon}). 
Before discussing entropy-related quantities, we present a more general result 
that allows one to obtain the hydrodynamic behavior of 
$\mathrm{Tr}({\mathcal F}(\Gamma^2_\ell))$ for arbitrary functions ${\mathcal F}(x)$. 
Our main result is that for any symbol of the form~\eqref{eq:symbolo} in the 
weakly-dissipative hydrodynamic limit $t,\ell\to\infty$ $\gamma\to0$ with $\gamma t$ and 
$t/\ell$ fixed, one has 
\begin{multline}
	\label{eq:tr-F-man}
	\mathrm{Tr}({\mathcal F}(\Gamma^2_\ell))=
	\int_{-\pi}^\pi\frac{dk}{2\pi}[2{\mathcal F}((A_k-C_k)^{2})
	-\mathrm{Tr}({\mathcal F}(\hat\Gamma_k^2))]\min(\ell,|2\varepsilon'_h(k)|t)\\
	+\ell\int_{-\pi}^\pi\frac{dk}{2\pi}\mathrm{Tr}({\mathcal F}(\hat\Gamma_k^{2})). 
\end{multline}
Here $\varepsilon'_h(k):=d\varepsilon_h(k)/dk$ (cf.~\eqref{eq:vareps}) 
is the group velocity of the Bogoliubov modes that diagonalize the 
Ising chain. This means that in the weak-dissipation limit the velocity 
of the quasiparticles is not affected by the dissipation. 
This is not the case for their correlation content, which can be quantified through the  functions $A_k,C_k$  
defined in~\eqref{eq:symbolo}. Eq.~\eqref{eq:tr-F-man}  
depends on the function $B_k$ via the symbol $\hat \Gamma_k$. 
The proof of~\eqref{eq:tr-F-man} relies on the 
multidimensional stationary phase approximation, and it is reported in~\ref{sec:useful-1} 
and~\ref{sec:useful-2}. Formula~\eqref{eq:tr-F-man} generalizes a result presented in 
Ref.~\cite{calabrese2012quantum} for $C_k=0$. Notice that since $C_k$ is an odd function and 
the integration domain in~\eqref{eq:tr-F-man} is symmetric around $k=0$, the 
minus sign in the term $A_k-C_k$ is irrelevant. Also, the restriction to 
functions ${\mathcal F}(\Gamma_\ell^2)$ of $\Gamma_\ell^2$ is not a 
severe limitation because the trace of the odd powers of $\Gamma_\ell$ vanishes 
by construction. 
Let us now discuss entropy-related quantities. The hydrodynamic behavior 
of the R\'enyi entropies $S^{\scriptscriptstyle(n)}$ is obtained by choosing 
\begin{equation}
	\label{eq:F-renyi}
	{\mathcal F}^{(n)}(z)=\frac{1}{2}\frac{1}{1-n}\ln\Big[\Big(\frac{1+\sqrt{z}}{2}\Big)^n
	+\Big(\frac{1-\sqrt{z}}{2}\Big)^n\Big]. 
\end{equation}
The von Neumann entropy corresponds to
\begin{equation}
	\label{eq:F-vn}
	{\mathcal F}(z)=-\frac{1}{2}\Big[\frac{1-\sqrt{z}}{2}\ln\Big(\frac{1-\sqrt{z}}{2}\Big)
	+\frac{1+\sqrt{z}}{2}\ln\Big(\frac{1+\sqrt{z}}{2}\Big)\Big].
\end{equation}
The correctness of~\eqref{eq:F-renyi} can be verified by 
checking that the Taylor series of ${\mathcal F}^{\scriptscriptstyle(n)}
(\Gamma_\ell^2)$ is consistent with~\eqref{eq:renyi-nu}. 
Thus, for a generic R\'enyi entropy 
$S_A^{\scriptscriptstyle(n)}$, by using~\eqref{eq:F-renyi} and~\eqref{eq:F-vn} 
we can rewrite~\eqref{eq:tr-F-man} as 
\begin{equation}
	\label{eq:f-entropy}
	S_A^{(n)}=
	\int_{-\pi}^\pi\frac{dk}{2\pi}\Big[s_k^{(n),YY}-s_k^{(n),\mathrm{mix}}\Big]\min(\ell,|2\varepsilon'_h(k)|t)
	+\ell\int_{-\pi}^\pi\frac{dk}{2\pi}s^{(n),\mathrm{mix}}_k. 
\end{equation}
Formula~\eqref{eq:f-entropy} has been conjectured recently in Ref.~\cite{carollo2021emergent}, 
and it has been shown to be valid for both free fermionic and free bosonic systems. 
In~\eqref{eq:f-entropy} the Yang-Yang entropies $s_k^{\scriptscriptstyle(n),YY}$ are given as  
\begin{equation}
	\label{eq:yy-n}
	s_k^{(n),YY}:=\frac{1}{1-n}\ln\Big[\rho_k^n+(1-\rho_k)^n\Big]. 
\end{equation}
In the limit $n\to1$ one obtains 
\begin{equation}
	\label{eq:yy-1}
	s^{YY}_k:=-\rho_k\ln(\rho_k)-(1-\rho_k)\ln(1-\rho_k).
\end{equation}
Here, in analogy with the case without dissipation (see, for instance, 
Ref.~\cite{alba2021generalized}), we defined a quasiparticle density 
$\rho_k$ as 
\begin{equation}
	\label{eq:rho-gen}
	\rho_k:=\frac{1-C_k+A_k}{2}. 
\end{equation}
Unlike the case without dissipation, now $\rho_k$ is time-dependent. 
Also, for $C_k\ne0$, $\rho_k$ is not even as a function of $k$, in contrast 
with the unitary case. 
The contribution $s^{\scriptscriptstyle (n),\mathrm{mix}}$ is defined as 
\begin{equation}
	\label{eq:renyi-full}
	s_k^{(n),\mathrm{mix}}=\frac{1}{2}\frac{1}{1-n}
	\mathrm{Tr}\ln\Big[\Big(\frac{\mathds{1}_2+\hat\Gamma_k}{2}\Big)^n+
\Big(\frac{\mathds{1}_2-\hat\Gamma_k}{2}\Big)^n\Big], 
\end{equation}
where $\hat\Gamma_k$ is the symbol of the Majorana 
correlator~\eqref{eq:gamma-k-inv}. 
Notice that~\eqref{eq:renyi-full} represents the contribution of the quasiparticle 
with quasimomentum $k$ to the R\'enyi entropy of the 
full system, which is zero in the case without dissipation.

Eq.~\eqref{eq:f-entropy} admits a simple physical interpretation. 
The first term in~\eqref{eq:f-entropy} describes the contribution of 
correlated quasiparticles to the entropies. A similar term appears in 
the quasiparticle picture in the absence of dissipation~\cite{alba2021spreading}. 
The term $s_k^{\scriptscriptstyle(n),\mathrm{mix}}$ (note 
the minus sign) in the square brackets encodes the fact that the dynamics 
is non unitary, and diminishes the total correlation between quasiparticles. 
In particular at $t\to\infty$ one has (see section~\ref{sec:ss-vanishing}) 
$s_k^{\scriptscriptstyle(n),\mathrm{mix}}\to s_k^{\scriptscriptstyle (n),\mathrm{YY}}$. 
The same contribution appears in the last term in~\eqref{eq:f-entropy}, which 
reflects the incoherent action of the environment. 
It is enlightening to consider the R\'enyi mutual information 
${\mathcal I}_{A:\bar A}^{\scriptscriptstyle(n)}$ between $A$ and $\bar A$. 
Due to the underlying quasiparticle picture that we have derived, 
the mutual information is expected to solely depend on the first 
term in~\eqref{eq:f-entropy}, i.e., the one describing the 
propagation of correlations through the dynamics of quasiparticles. One thus expects 
\begin{equation}
	\label{eq:mi-quasi}
	{\mathcal I}^{(n)}_{A:\bar A}=2\int \frac{dk}{2\pi}\Big[s_k^{(n),YY}-
	s_k^{(n), \mathrm{mix}}\Big]\min(\ell,2|\varepsilon_h'(k)|t), 
\end{equation}
which assumes that the mutual information is only sensitive to 
the contribution of the correlated quasiparticles shared by the bipartition. 
The validity of such  assumption has been thoroughly verified numerically. Interestingly, this suggests 
that although ${\mathcal I}^{\scriptscriptstyle(n)}_{A:\bar A}$ is not a proper 
entanglement measure for mixed states, it shares some expected features of a proper 
entanglement measure. An important remark is that in deriving~\eqref{eq:mi-quasi} 
we assumed $\ell\ll L$, and $t\ll L-\ell$ to avoid the effect of revivals at long 
times. Within the quasiparticle picture these are due to quasiparticles reentering 
the subsystem. These effects can, in principle, be incorporated in~\eqref{eq:f-entropy}, 
similarly to the case without dissipation~\cite{modak2020entanglement}. 

\subsection{Quasiparticle densities and the case of even dissipation}
\label{sec:quasi-rho}

The quasiparticle density $\rho_k$ (cf.~\eqref{eq:rho-gen}) is a key ingredient 
in~\eqref{eq:f-entropy} and~\eqref{eq:mi-quasi}. From~\eqref{eq:gamma-1}, 
\eqref{eq:gamma-2} and~\eqref{eq:rho-gen} one obtains that 
\begin{equation}
	\label{eq:rho-k-fin}
	\rho_k=\frac{1}{2}-\frac{1}{2}e^{-4\gamma(s_\mathrm{e}+z_\mathrm{o})t}\cos(\Delta_k)
	-\frac{1}{2}\frac{s_\mathrm{o}-
	z_\mathrm{e}}{s_\mathrm{e}+z_\mathrm{o}}(1-e^{-4\gamma(s_\mathrm{e}+z_\mathrm{o})t}). 
\end{equation}
Clearly, $\rho_k$ is time-dependent, as anticipated in the previous 
sections. 
From~\eqref{eq:rho-k-fin}, it is straightforward to verify that 
$\rho_k$ satisfies the simple rate equation as 
\begin{equation}
	\label{eq:rate-eq}
	\frac{d\rho_k}{dt}=-4\gamma(s_\mathrm{e}+z_\mathrm{o})\rho_k+
	2\gamma(s_\mathrm{e}-s_\mathrm{o}+z_\mathrm{o}+z_\mathrm{e}). 
\end{equation}
For $z_\mathrm{o}=s_\mathrm{o}=0$, Eq.~\eqref{eq:rate-eq} has been discussed 
in Ref.~\cite{carollo2021emergent}. 
In the absence of dissipation, one has 
$z_{\mathrm{e}}=z_\mathrm{o}=s_\mathrm{e}=s_{\mathrm{o}}=0$ 
and $\rho_k$ becomes $\rho_k=(1-\cos(\Delta_k))/2$, i.e., it is conserved 
during the dynamics. 
Interestingly, due to the terms $z_\mathrm{o}$ and $s_{\mathrm{o}}$, 
$\rho_k$ is not an even function of $k$. The condition $z_\mathrm{o}=s_\mathrm{o}=0$ 
implies that $C_k=0$ (cf.~\eqref{eq:symbolo}). For this reason, 
here we define this dissipation 
as \emph{even}  dissipation. 
This type of dissipation has been considered recently in Ref.~\cite{carollo2021emergent}. 
Remarkably, for even dissipation, $\rho_k$ coincides with the density of the Bogoliubov 
modes $\alpha_k$ that diagonalize the Ising chain (see section~\ref{sec:model}). 
This can be verified by comparing~\eqref{eq:rho-k-fin} and~\eqref{eq:rho-gamma} 
after using that  the symbol $\hat\Gamma_k$ for even dissipation reads as 
\begin{equation}
	\label{eq:symb-off-d}
	\hat\Gamma_k=-e^{-4\gamma s_\mathrm{e}t}(\cosh(\Delta_k)\sigma_y^{(k)}
	+\sinh(\Delta_k)\sigma_x^{(k)}e^{2i\varepsilon_h(k)t\sigma_y^{(k)}})+
	\frac{z_\mathrm{e}}{s_\mathrm{e}}(1-e^{-4\gamma s_\mathrm{e}t})\sigma_y^{(k)}. 
\end{equation}
Interestingly,  Eq.~\eqref{eq:symb-off-d} has the same structure 
as for the quench without  dissipation~\cite{calabrese2012quantum},  
although there the functions $A_k,B_k$ are different and do not depend on time. 
It is also worth remarking that the dissipation with gain/loss processes discussed 
in section~\ref{sec:lin} is a simple example of even dissipation, and it 
corresponds to the choice $s_\mathrm{e}=g^++g^-$ and 
$z_\mathrm{e}=(g^--g^+)\cos(\theta_k)$.

On the other hand, for generic dissipation, $\rho_k$ is not the density of 
the Bogoliubov modes $\alpha_k$ (cf.~\eqref{eq:free}) of the quantum Ising chain. 
Indeed, the dynamics of $\alpha_k$ (cf.~\eqref{eq:free}) is 
obtained by using~\eqref{eq:rho-gamma}, which depends only on the off-diagonal 
matrix elements of $\hat\Gamma_k$, whereas in general~\eqref{eq:rho-k-fin} depends on 
the full matrix $\hat\Gamma_k$.

\subsection{Vanishing of the mutual information in the steady state }
\label{sec:ss-vanishing}

It is interesting to investigate the behavior of the mutual information 
in the long time limit $\gamma t\to\infty$. 
For physical choices of the functions $a(k),b(k),c(k)$, 
in the limit $\gamma t\to\infty$ the steady-state density $\rho_k(\infty)$ is 
obtained from~\eqref{eq:rho-k-fin} as 
\begin{equation}
	\label{eq:ss-density}
\rho_k(\infty)=\frac{1}{2}\Big(1-\frac{s_o-z_e}{s_e+z_o}\Big). 
\end{equation}
Eq.~\eqref{eq:ss-density} allows us to obtain the  Yang-Yang 
 contribution to the steady-state entropy 
(the term $s_k^{\scriptscriptstyle(n),\mathrm{YY}}$ in~\eqref{eq:f-entropy}). 
Let us now discuss the term $s_k^{\scriptscriptstyle(n),\mathrm{mix}}$ due to mixedness of the quantum state 
in~\eqref{eq:f-entropy}. We get 
\begin{align}
	\label{eq:asy-1}
	& \hat\Gamma_k^{(1)}\xrightarrow{\gamma t\to\infty} 0\\
	\label{eq:asy-2}
	& \hat\Gamma_k^{(2)}\xrightarrow{\gamma t\to\infty} 
	\frac{1}{s_\mathrm{e}^2-z_\mathrm{o}^2}\Big[
	(s_\mathrm{e}s_\mathrm{o}+z_\mathrm{e}z_\mathrm{o})\mathds{1}_2
+(s_\mathrm{e}z_\mathrm{e}+s_\mathrm{o}z_\mathrm{o})\sigma_y^{(k)}\Big]. 
\end{align}
It is now straightforward to derive the eigenvalues $\nu_{\pm}$ of 
$\hat\Gamma_k=\hat\Gamma_k^{\scriptscriptstyle(1)}+\hat\Gamma_k^{\scriptscriptstyle(2)}$  
as
\begin{equation}
	\label{eq:nu}
	\nu_\pm=\frac{s_\mathrm{o}\pm z_\mathrm{e}}{s_\mathrm{e}\mp z_\mathrm{o}}. 
\end{equation}
By using~\eqref{eq:nu} and~\eqref{eq:ss-density} in~\eqref{eq:mi-quasi}, one can verify 
that in the limit $\gamma t\to\infty$, $s^{YY}_k-s^\mathrm{mix}_k$ is 
an odd function of $k$, which implies that its integral vanishes in~\eqref{eq:f-entropy}. 
We note that the fact that for $\gamma t\to\infty$ $s_k^{\scriptscriptstyle (n),\mathrm{YY}}-
s_k^{\scriptscriptstyle (n),\mathrm{mix}}$ is an odd function of  $k$ might sound 
troubling at first, since it seems to suggest that one of the quasiparticles 
carries a negative correlation content. However, we observe that when considering the correlation shared by pairs of quasiparticles with momenta $\pm k$, only the total correlation $s_k^{\scriptscriptstyle (n),\mathrm{YY}}-
s_k^{\scriptscriptstyle (n),\mathrm{mix}}+s_{-k}^{\scriptscriptstyle (n),\mathrm{YY}}-
s_{-k}^{\scriptscriptstyle (n),\mathrm{mix}}$ makes physical sense. In particular, this suggests the rewriting of Eq.~\eqref{eq:f-entropy} as follows
\begin{equation}
	\label{eq:f-entropy_tot}
	\begin{split}
	S_A^{(n)}&=
	\int_0^\pi \frac{dk}{2\pi}\Big[s_k^{(n),YY}+s_{-k}^{(n),YY}-s_k^{(n),\mathrm{mix}}-s_{-k}^{(n),\mathrm{mix}}\Big]\min(\ell,|2\varepsilon'_h(k)|t)\\
	&+\ell\int_0^\pi\frac{dk}{2\pi}\left(s^{(n),\mathrm{mix}}_k+s^{(n),\mathrm{mix}}_{-k}\right),
	\end{split}
\end{equation}
which clearly shows that the total correlation between quasiparticles is actually an even function of $k$. In the stationary state this correlation content is thus zero for every quasiparticle pair. 
We note that similar considerations on the total correlations between pairs apply also in the absence 
of dissipation for quenches from inhomogeneous initial states~\cite{alba2018entanglementand,alba2019entanglement,alba2019towards,mestyan2020molecular}. 

\section{Numerical benchmarks}
\label{sec:numerical}

We now provide numerical benchmarks of the results derived 
in section~\ref{sec:result}. We focus on the subsystem entropy 
in section~\ref{sec:enta}. In section~\ref{sec:mi-numerics} and 
section~\ref{sec:unp} we discuss the behavior of the mutual information for 
the most general linear dissipation. 

\subsection{Subsystem entropy}
\label{sec:enta}

\begin{figure}[t]
\begin{center}
\includegraphics[width=.5\textwidth]{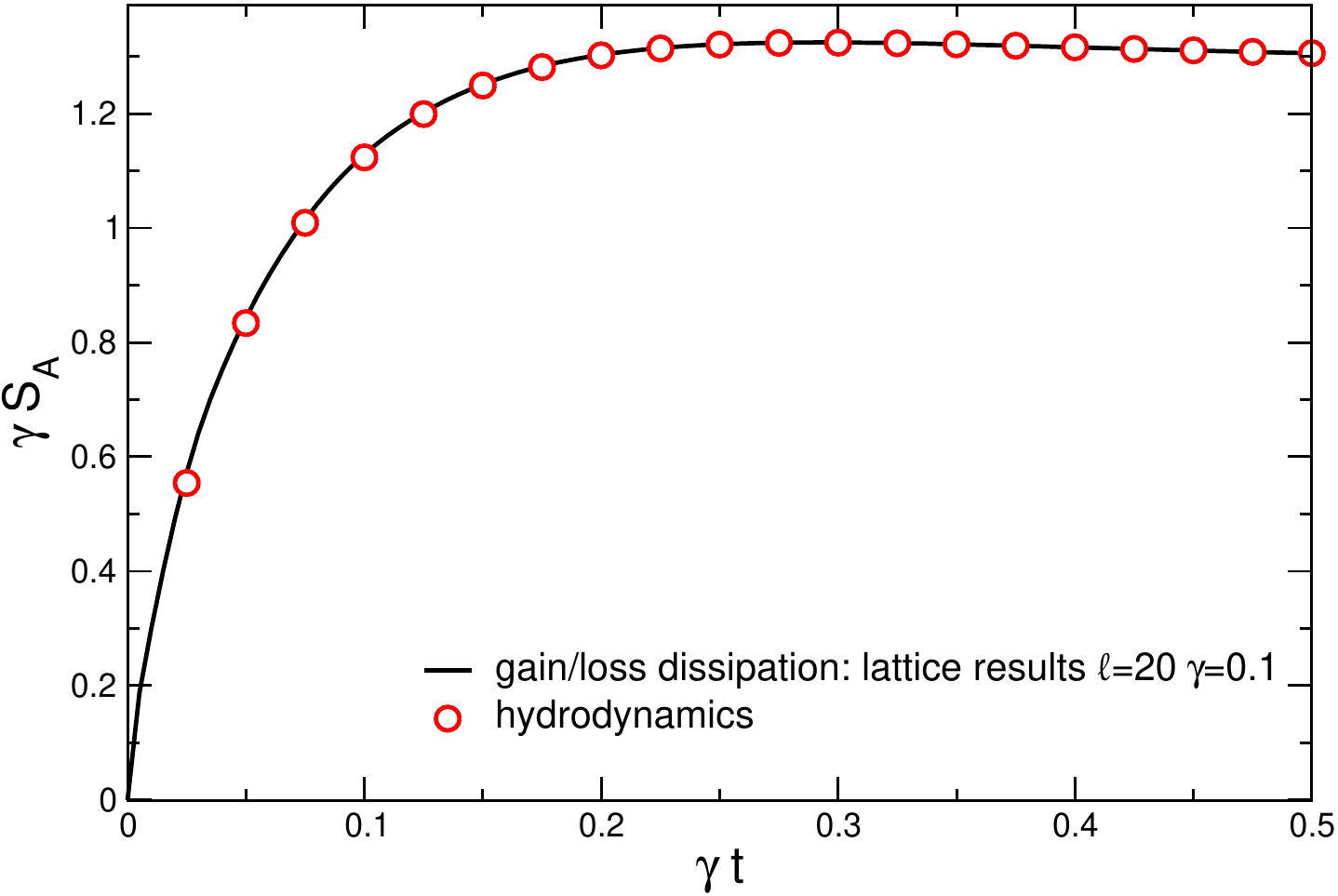}
\caption{ Dynamics of the subsystem von Neumann entropy $S_A$ 
 in the Ising chain after a magnetic field quench $h_0\to h$ 
 ($h_0=0.1,h=2$) in the presence of diagonal gain/loss dissipation. 
 The dissipation corresponds to Kossakowski matrix with 
 $a=b=3/2$ $c=-1/2i$ (cf.~\eqref{eq:K-symb}), 
 and $\gamma=0.1$. The figure shows $\gamma S_A$ versus 
 $\gamma t$. The continuous line denotes exact lattice results. 
 The circles are the theory predictions in the weakly-dissipative 
 hydrodynamic limit $\gamma\to0$, $t,\ell\to\infty$ with 
 $\gamma\ell$ and $t/\ell$ fixed. 
}
\label{fig1:entropy}
\end{center}
\end{figure}

Here we consdier the entropy $S_A$ of a finite subsystem $A$ of 
length $\ell$ embedded in an infinite chain (see Fig.~\ref{fig0:cartoon}). 
In Fig.~\ref{fig1:entropy} we show 
numerical data for $S_A$. We consider the quench 
from the ground state of the Ising chain with initial magnetic field $h_0=0.1$ and 
final one $h=2$. We focus on gain/loss dissipation (see section~\ref{sec:lin}) 
with gain rate $\gamma g^+$ and loss rate $\gamma g^-$. We fix $g^+=0.5$ and $g^-=1$. 
This corresponds to Kossakowski matrix~\eqref{eq:K-symb} with $a=b=3/2$ and $c=-1/2i$  
This gives $z_\mathrm{o}=s_{\mathrm{o}}=0$ (cf.~\eqref{eq:so-def} and~\eqref{eq:z-def}), implying 
that, as discussed already, gain/loss dissipation is a particular case of 
the even dissipation discussed in section~\ref{sec:quasi-rho}. We should mention that  
dissipation with non-local losses~\cite{carollo2021emergent} is even as well. 
In Fig.~\ref{fig1:entropy} we fix the strength of the dissipation 
$\gamma=0.1$. The continuous black line denotes numerical exact results obtained by using 
the analytic expression for the time-evolved correlation matrix $\Gamma$ (cf.~\eqref{eq:gamma-def}) 
and the results in~\ref{sec:obs}. The circles in the Figure are the analytic 
results in the weakly-dissipative hydrodynamic limit (cf.~\eqref{eq:f-entropy}). 
Although Eq.~\eqref{eq:f-entropy} 
is expected to hold in the limit $\gamma\to0$ the agreement between the numerics and the 
analytic result is remarkable. 

Notice that $\gamma S_A$ attains a finite value at $\gamma t\to\infty$. This is 
due to the second term in~\eqref{eq:f-entropy}, which is sensitive only to the dissipative 
processes. On the other hand, the first term in~\eqref{eq:f-entropy} describes the 
contribution to $S_A$ of correlated pairs of quasiparticles, similar to the 
case without dissipation. As it was discussed in section~\ref{sec:ss-vanishing} 
this vanishes at long times. To extract these contributions, it is convenient to 
focus on the mutual information, as it is clear from~\eqref{eq:mi-quasi}.

\begin{figure}[t]
\begin{center}
\includegraphics[width=.5\textwidth]{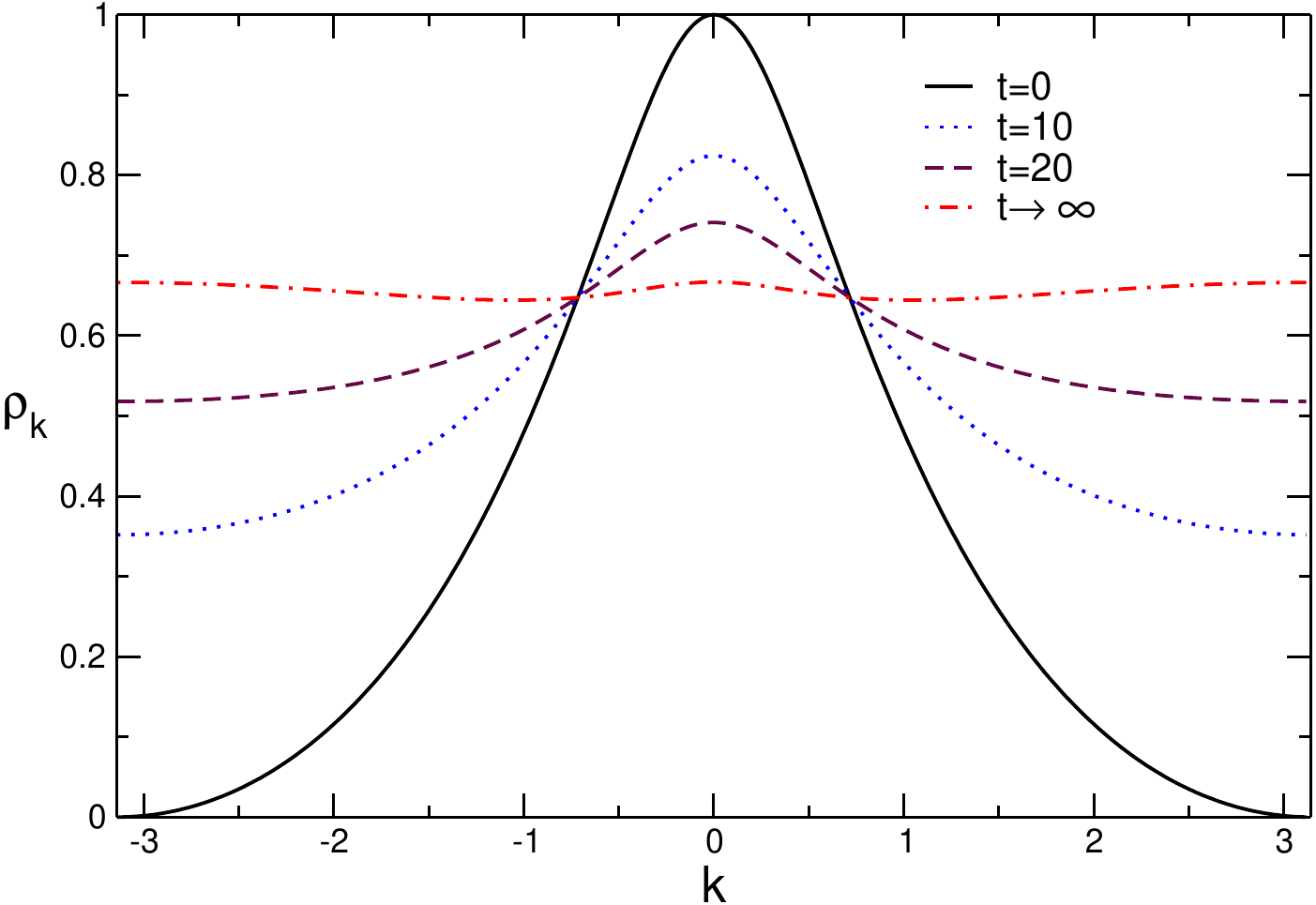}
\caption{Evolution of quasiparticle densities in the Ising chain 
 with linear dissipation after a magnetic field quench $h_0\to h$. 
 All the results are for the $h_0=0.1$ and $h=2$. 
 Here we consider diagonal gain/loss dissipation, which corresponds to Kossakowski 
 matrix (cf.~\eqref{eq:K-symb}) with $a=b=3/2$ and 
 $c=-0.5i$, and $\gamma=0.0125$. $\rho_k$ is obtained from~\eqref{eq:rho-gen}. 
 The red dashed-dotted line is the result for $t\to\infty$. 
}
\label{fig4a:density}
\end{center}
\end{figure}

\begin{figure}[t]
\begin{center}
\includegraphics[width=.5\textwidth]{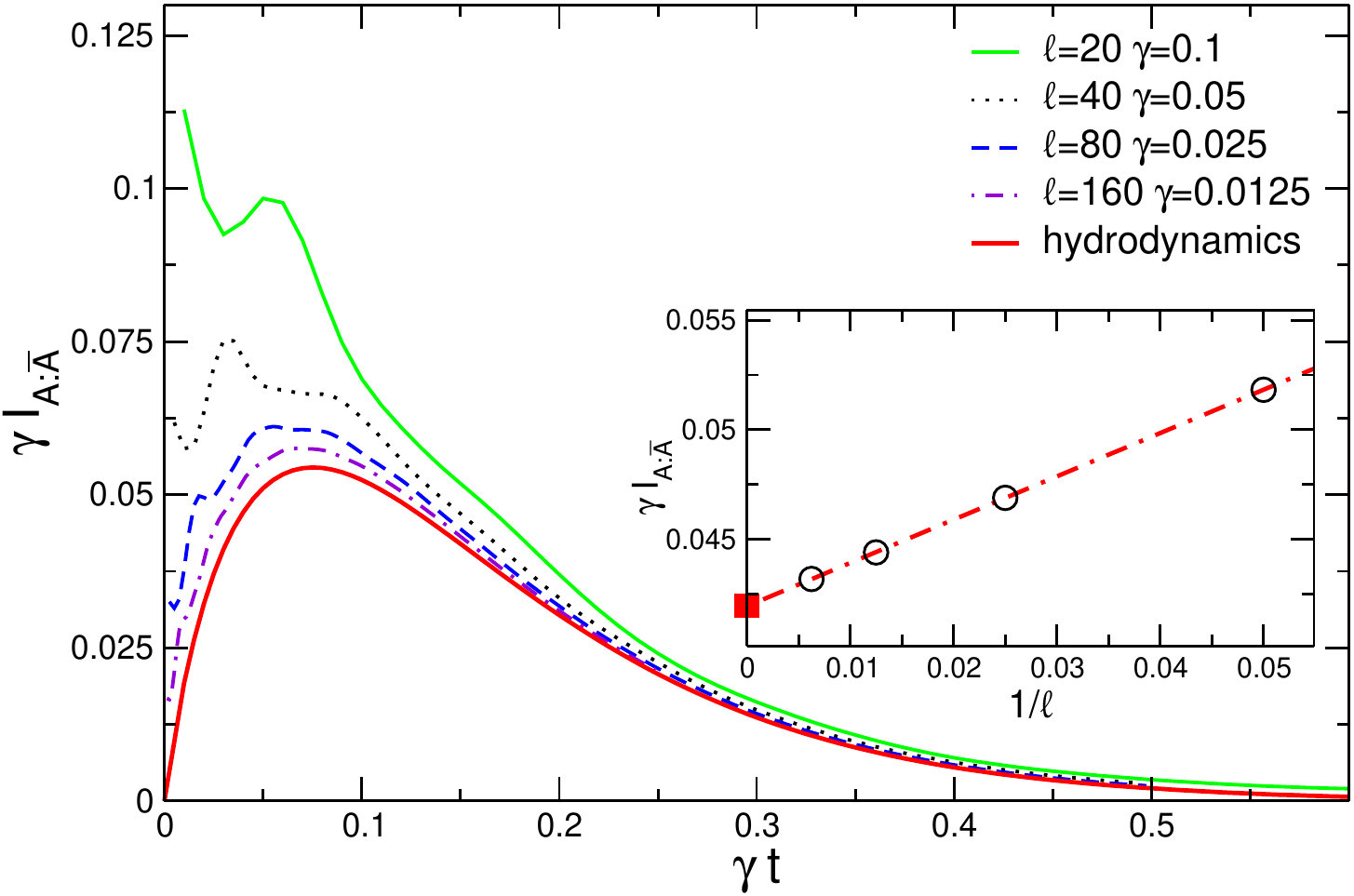}
\caption{Dynamics of the mutual information ${\mathcal I}_{A:\bar A}$ in the Ising chain 
 with diagonal gain/loss dissipation after the magnetic field quench $h_0\to h$ ($h_0=0.1,h=2$). 
 The dissipation corresponds to Kossakowski matrix (cf.~\eqref{eq:K-symb}) 
 with $a=b=3/2$ $c=-1/2i$ and $\gamma=1/(2\ell)$. In the main figure we 
 plot $\gamma {\mathcal I}_{A:\bar A}$ versus $\gamma t$. The different lines 
 are exact lattice results for different $\ell$. The continuous red line is the 
 expected result in the weakly-dissipative hydrodynamic limit $\ell,t\to\infty$ 
 with $\gamma\ell$ $\gamma t$ fixed. Scaling corrections due to finite $t,\ell$ are 
 present. Inset: Finite-size scaling analysis. We plot $\gamma {\mathcal I}_{A:\bar A}$ 
 versus $1/\ell$ at fixed $\gamma t=0.15$. The circles are the same data as in the 
 main Figure, the square symbol at $\ell\to\infty$ is the hydrodynamic limit result. 
 The dashed-dotted line is a linear fit. 
}
\label{fig2:gl-mi}
\end{center}
\end{figure}

\subsection{Mutual information}
\label{sec:mi-numerics}

Here we focus on the mutual information ${\mathcal I}_{A:\bar A}$ between 
interval $A$ and its complement. As it is clear from~\eqref{eq:mi-quasi}, 
the mutual information is solely sensitive to the correlated pairs that are 
produced after the quench and shared by the bipartition. Specifically, 
in constructing the mutual information the second term 
in~\eqref{eq:f-entropy} cancels out. However, the mutual information 
does not represent a proper measure of entanglement since 
quasiparticles, in these cases, are both quantum and classically correlated. 

We first consider the case of gain/loss dissipation, as in section~\ref{sec:enta}. 
Before discussing the mutual information it is useful to consider the quasiparticle 
density $\rho_k$, which  determines (cf.~\eqref{eq:mi-quasi}) the 
dynamics of the mutual information. We plot $\rho_k$ in Fig.~\ref{fig4a:density} versus 
the quasimomentum $k$, for several times after the quench. At $t=0$ one has 
$\rho_k=(1-\cos(\Delta_k))/2$. The initial $\rho_k$ exhibits a maximum at 
$k=0$ and it vanishes at $\pm\pi$. At long times $\gamma t\to\infty$, larger 
momenta get populated, although $\rho_k$ is not completely flat in momentum space. 
The steay-state $\rho_k$ is obtained from~\eqref{eq:ss-density}. 
For gain and loss dissipation 
with rates $\gamma g^+$ and $\gamma g^-$ 
one has $z_\mathrm{o}=s_\mathrm{o}=0$ and 
$z_\mathrm{e}=-(g^--g^+)\cos(\theta_k)$ 
(cf.~\eqref{eq:z-def}~\eqref{eq:ze-def}~\eqref{eq:theta}), which, together with~\eqref{eq:ss-density}, imply  that 
$\rho_k(\infty)=(1-z_\mathrm{e}/s_\mathrm{e})/2$. Notice that $\rho_k$ is an even 
function of $k$, as expected for even dissipation. 

We show numerical results for the mutual information in 
Fig.~\ref{fig2:gl-mi}. We consider the same parameters as in Fig.~\ref{fig1:entropy}, 
plotting the rescaled von Neumann mutual information 
$\gamma{\mathcal I}_{A:\bar A}$ versus rescaled time $\gamma t$. The 
strength of the dissipation $\gamma$ is rescaled as $2/\ell$. 
The red continuous line is the analytic result in the weakly-dissipative 
hydrodynamic limit (cf.~\eqref{eq:mi-quasi}). 
In contrast with the results for the entropy 
$S_A$ (see Fig.~\ref{fig1:entropy}), the data for 
the mutual information exhibit sizeable corrections, with oscillating behavior. At 
large $\gamma t$ the mutual information decays, as predicted by~\eqref{eq:mi-quasi}.  
Eq.~\eqref{eq:mi-quasi} is valid only in the weakly-dissipative 
hydrodynamic limit. Indeed, upon increasing $\ell$ and decreasing $\gamma$, 
the numerical data approach the theory predictions. This is checked in the inset 
of Fig.~\ref{fig2:gl-mi}, showing 
$\gamma{\mathcal I}_A$ versus $1/\ell$ at fixed $\gamma t=0.15$. The dashed-dotted line 
is a fit to a $1/\ell$ behavior, whereas the full square symbol is the result for 
$\ell\to\infty$. The agreement with~\eqref{eq:mi-quasi} is remarkable. We should mention 
that similar scaling corrections as $1/\ell$ are present in the case without 
dissipation~\cite{alba2017entanglement}. 

To provide a more stringent check of the results of section~\ref{sec:result} we 
now consider a more complicated dissipation. Specifically, we choose the  Kossakowski 
matrix (cf.~\eqref{eq:Ko-k}) with parameters $c=\sin(k)-1/2i$, $a=b=3/2$. Although 
the dissipation is non-diagonal, it is still even (see section~\ref{sec:quasi-rho}).

\begin{figure}[t]
\begin{center}
\includegraphics[width=.5\textwidth]{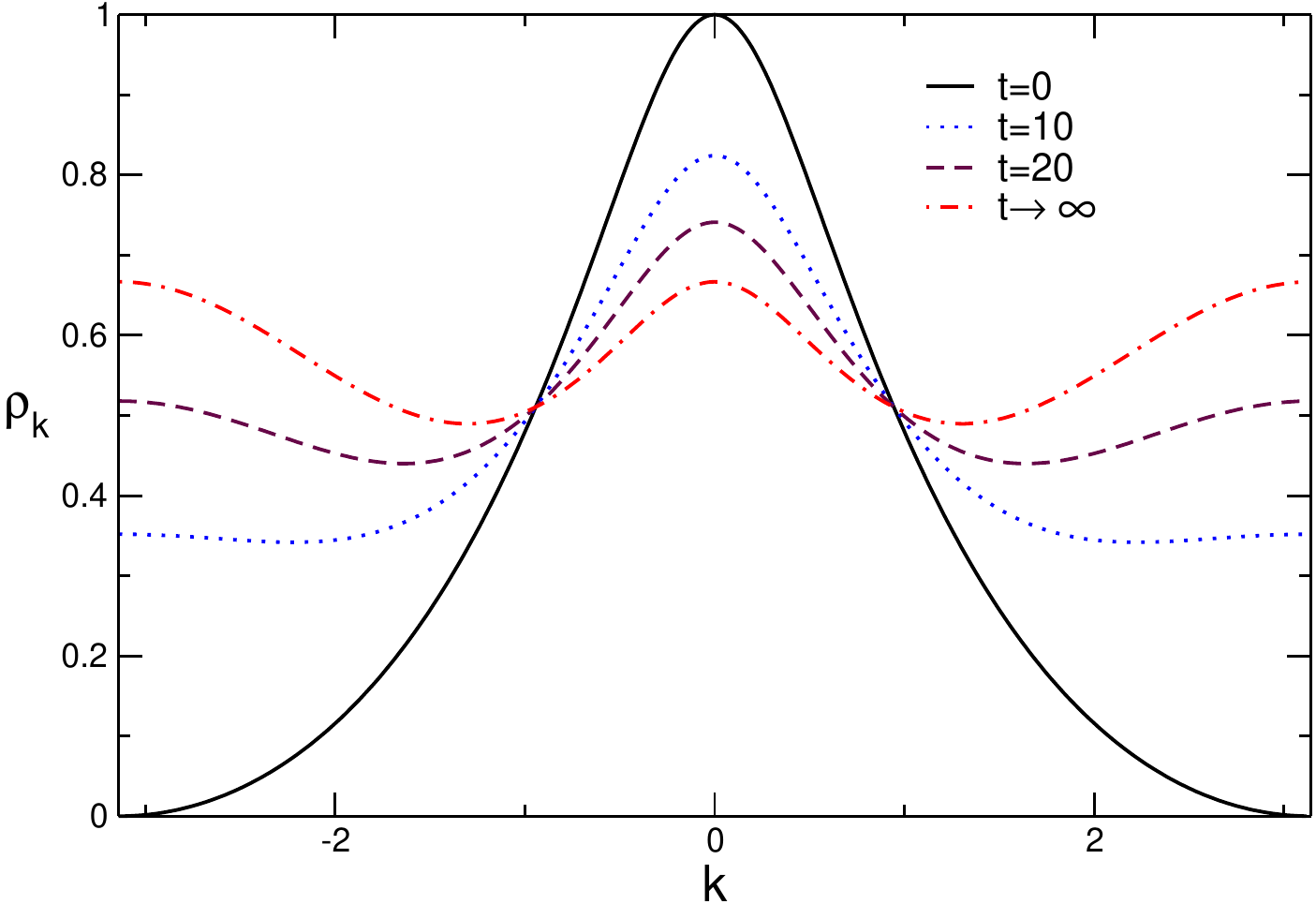}
\caption{Evolution of quasiparticle densities $\rho_k$ in the Ising chain 
 with linear dissipation after a magnetic field quench $h_0\to h$. 
 All the results are for the $h_0=0.1$ and $h=2$. 
 The Kossakowski matrix (cf.~\eqref{eq:Ko-k}) 
 encoding dissipation has parameters as 
 $a=b=3/2$, $c=\sin(k)-1/2i$, and $\gamma=0.0125$. 
 $\rho_k$ is obtained from~\eqref{eq:rho-gen}.  
 As in Fig.~\ref{fig2:gl-mi} the dissipation is even. 
 Indeed, one has $\rho_k=\rho_{-k}$. 
}
\label{fig4b:density}
\end{center}
\end{figure}

We first discuss the quasiparticles densities $\rho_k$ in Fig.~\ref{fig4b:density}. 
The behavior is qualitatively similar to that observed in Fig.~\ref{fig4a:density}. 
The intial density is peaked around $k=0$, and it vanishes at 
$k=\pm\pi$, whereas at long times quasiparticles with larger $k$ are populated. At $t\to\infty$ 
the density exhibits oscillating beahavior as a function of $k$, as in Fig.~\ref{fig4a:density}. 
%
\begin{figure}[t]
\begin{center}
\includegraphics[width=.5\textwidth]{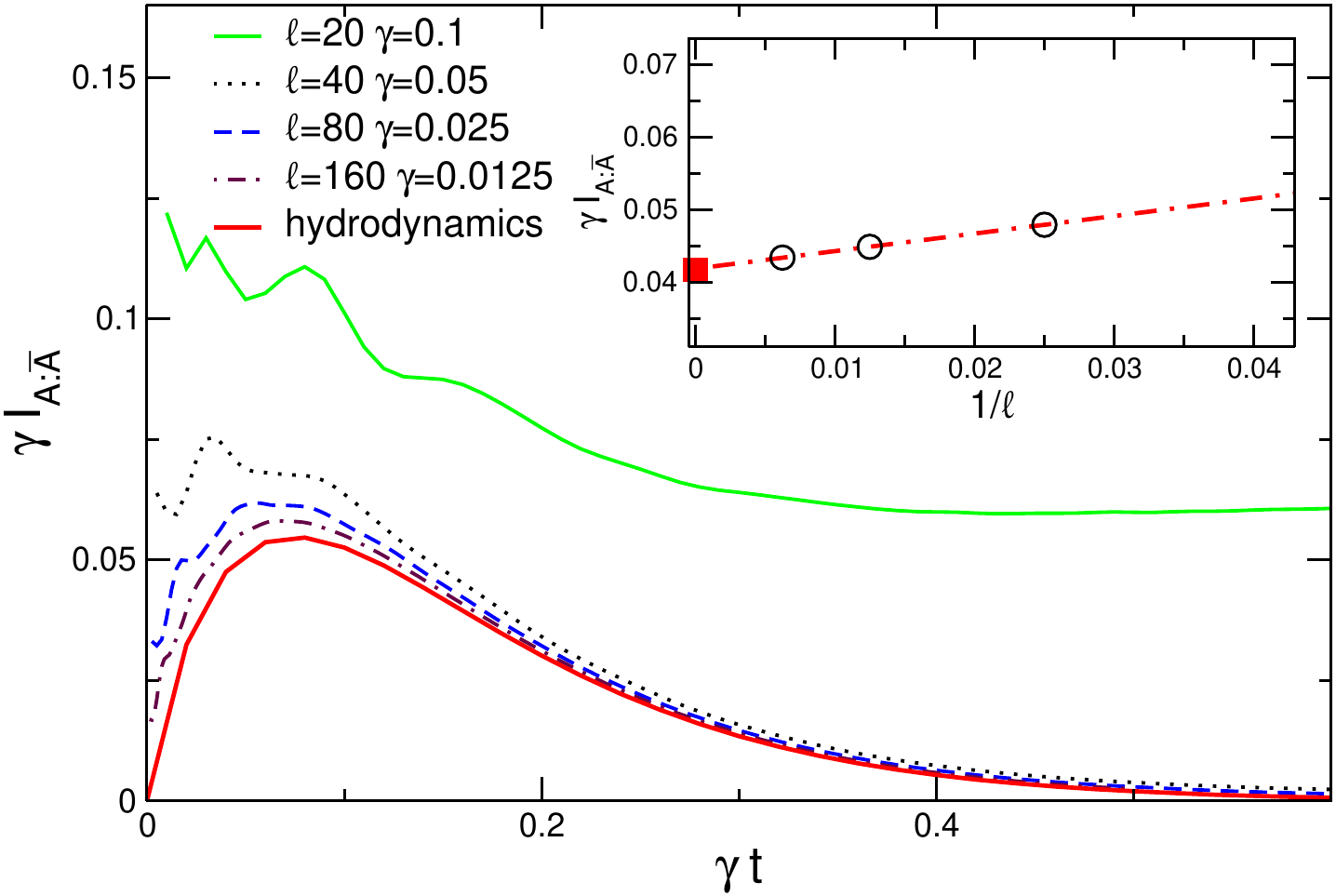}
\caption{The same as in Figure~\ref{fig3:gl-d2} for the dissipation given 
 by the Kossakowski matrix (cf.~\eqref{eq:K-symb}) with 
 $a=b=3/2$, $c=\sin(k)-1/2 i$, where $k$ is the quasimomentum. The inset 
 show the finite-size scaling analysis for fixed $\gamma t=0.15$. 
}
\label{fig4:gl-d3}
\end{center}
\end{figure}
%
The mutual information is reported in 
Fig.~\ref{fig4:gl-d3}. The 
data exhibit oscillating scaling corrections for small $\ell$ and short times. 
These corrections decay as $1/\ell$. 
In the hydrodynamic scaling limit the agreement between the numerical data and the 
quasiparticle picture is perfect, as shown in the inset. 
Notice that, similar to the gain/loss  dissipation (see Fig.~\ref{fig4:gl-d3}), the 
mutual information vanishes at $t\to\infty$, in agreement with the results of 
section~\ref{sec:ss-vanishing}.

\begin{figure}[t]
\begin{center}
\includegraphics[width=.5\textwidth]{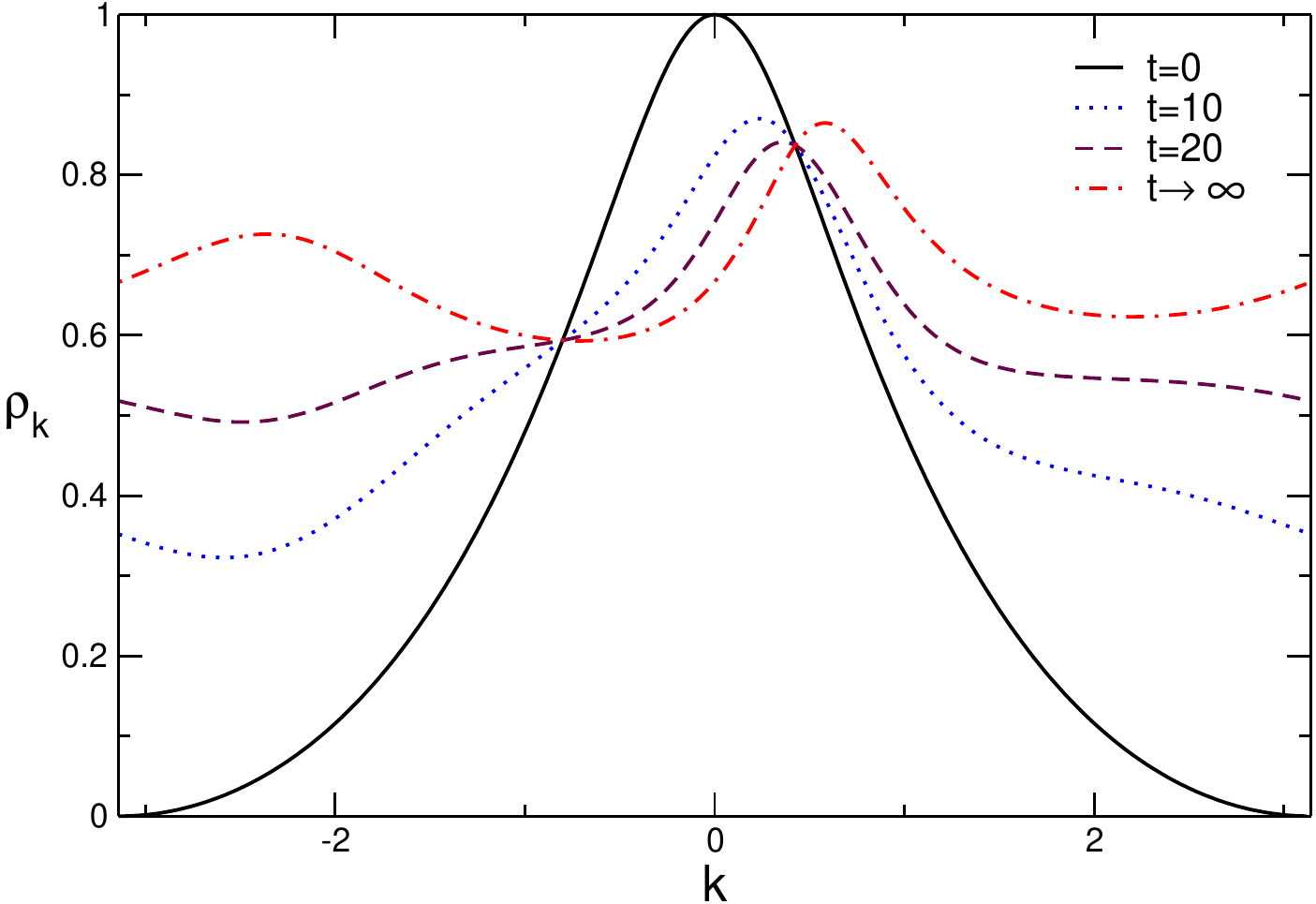}
\caption{Evolution of the effective quasiparticle densities 
 $\rho_k=(1-C_k+A_k)/2$ (cf.~\eqref{eq:rho-gen}) after a quantum quench 
 in the dissipative Ising chain. Results are for the quench with 
 $h_0=0.1$ and $h=2$. The dissipation corresponds to Kossakowski matrix 
 (cf~\eqref{eq:K-symb}) with  $a=b=3/2$, $c=5/2\cos(k)-1/2i$, 
 and $\gamma=0.0125$. Notice that $\rho_k\ne\rho_{-k}$. 
}
\label{fig5:density}
\end{center}
\end{figure}

\begin{figure}[t]
\begin{center}
\includegraphics[width=.5\textwidth]{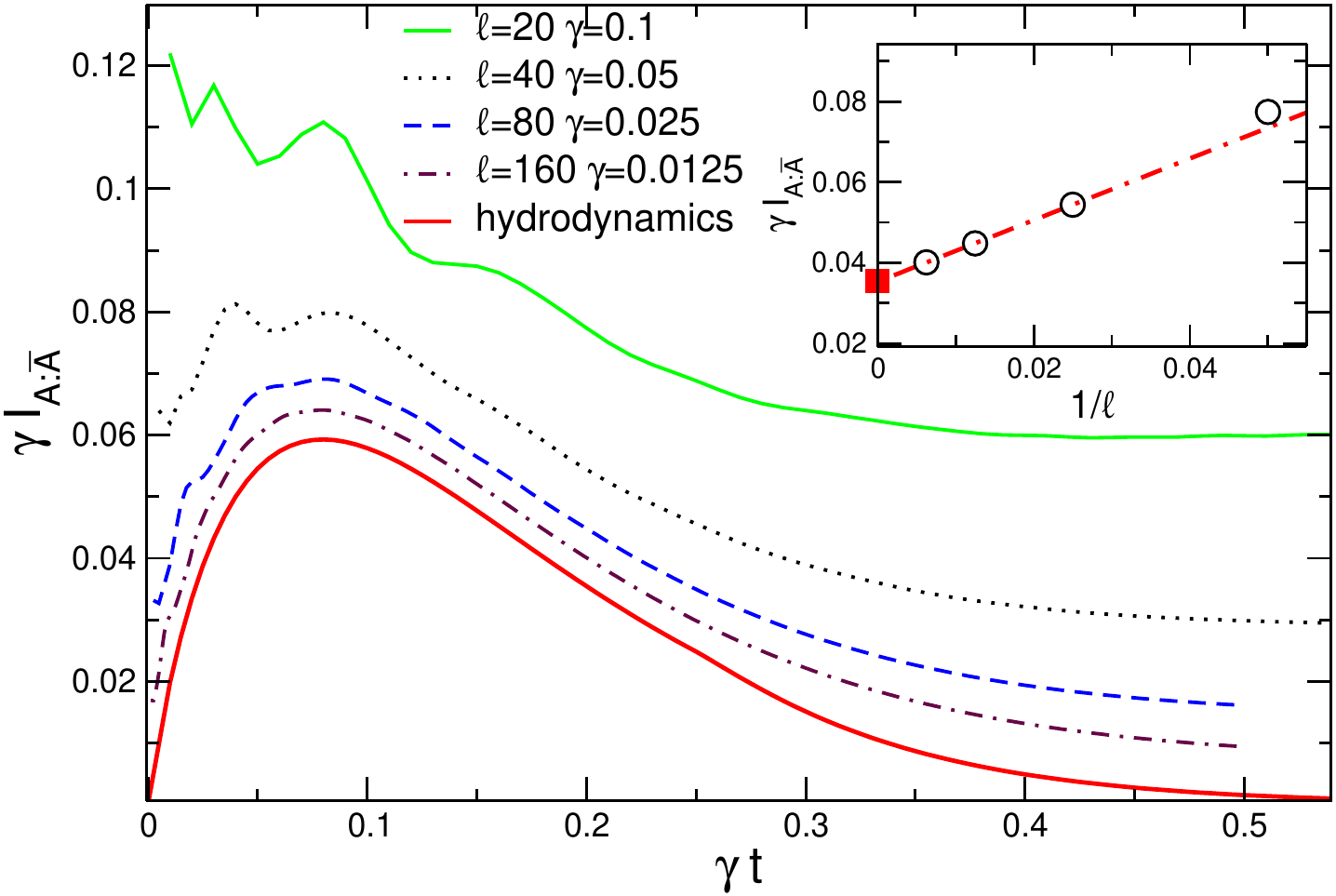}
\caption{Dynamics of the mutual information ${\mathcal I}_{A:\bar A}$ in the Ising chain 
 with linear dissipation after the magnetic field quench $h_0\to h$ ($h_0=0.1,h=2$). 
 The dissipation corresponds to the Kossakowski matrix (cf.~\eqref{eq:K-symb}) 
 with $a=b=3/2$, $c=5/2\cos(k)-1/2i$, with $k$ the quasimomentum. The dissipation 
 strength $\gamma$ is rescaled as $\gamma=1/(2\ell)$. In the main figure we 
 plot $\gamma {\mathcal I}_{A:\bar A}$ versus $\gamma t$. Different lines 
 are exact lattice results for different sizes $\ell$ of $A$. The continuous red line is the 
 result in the weakly-dissipative hydrodynamic limit. 
 Scaling corrections due to the finite $t,\ell$ are 
 present. Inset: Finite-size scaling analysis. We plot $\gamma {\mathcal I}_{A:\bar A}$ 
 versus $1/\ell$ at fixed $\gamma t=0.2$. The circles are the same data as in the 
 main Figure, the square symbol at $\ell\to\infty$ is the hydrodynamic limit result. 
 The dashed-dotted line is a linear fit. 
}
\label{fig3:gl-d2}
\end{center}
\end{figure}

\subsection{Evolution under an ``unphysical'' dissipation}
\label{sec:unp}

A crucial condition for the Lindblad equation~\eqref{eq:liouv} to be 
physical is that the Kossakowski matrix is positive semidefinite. 
This is necessary to ensure that the Lindblad evolution is a completely 
positive and trace-preserving map~\cite{petruccione2002the}. Moreover, 
any fermionic density matrix 
$\rho$ is characterized by its Majorana covariance matrix $\Gamma$ 
(cf.~\eqref{eq:gamma-def}). The matrix $\Gamma$ is by definition 
purely imaginary and antisymmetric, which implies that its eigenvalues 
are real and are arranged in pairs $\pm \nu_j$. The only condition for 
$\Gamma$ to correspond to a physical density matrix is that $|\nu_j|\le 1$. 
An interesting observation is that this condition 
can be satisfied even if the Kossakowski matrix is not positive semidefinite. 
This could mean that the dynamical map is positive, even if not completely 
positive, or that it is not even positive but still maps the initial 
state considered into a well-defined state. 
Even in these ``unphysical'' situations the analytical results derived in 
section~\ref{sec:result} hold true. 

Here, in order to show this, we consider the Kossakowski 
matrix with $a=b=3/2$ and $c=5/2\cos(k)-1/2i$. 
First, now one has that $C_k\ne0$ (cf.~\eqref{eq:symbolo}). This implies 
that we are not considering even dissipation. 
This is shown in Fig.~\ref{fig5:density}. As it is clear from the figure, 
$\rho_k$ is not an even function of $k$. 
One can check that the eigenvalues of $K$ are not positive. 
However,  we verified numerically that at 
any time $t$, the eigenvalues of $\Gamma$ satisfy the condition $|\nu_j|\le1$. 
 The validity of our formulae for this unphysical dissipation is verified 
in Fig.~\ref{fig3:gl-d2} focusing on the mutual information.    
Despite the fact that the dissipation is not even, and that the 
evolution is not a completely positive map, the qualitative 
behavior of ${\mathcal I}_{A:\bar A}$ 
is the same as for the other types of dissipation explored so far. 
Specifically, the mutual information exhibits a peak at intermediate times 
and it decays exponentially at $t\to\infty$. 
Notice, however, the large scaling corrections, which are 
discussed in the inset of Fig.~\ref{fig3:gl-d2} at fixed $\gamma t=0.2$.

\section{Conclusions}
\label{sec:concl}

We investigated the out-of-equilibrium dynamics after a generic 
magnetic field quench in the transverse field Ising chain in the presence of 
the most general linear dissipation that can be treated within the 
framework of Markovian master equations~\cite{prosen2008third}. 
Our main result is formula~\eqref{eq:tr-F-man}, 
which provides an analytic expression for the dynamics of any function of the Majorana 
covariance matrix, in the weakly-dissipative hydrodynamic limit. 
By using~\eqref{eq:tr-F-man} we derived exact results for the 
dynamics of von Neumann and R\'enyi entropies, and of the associated mutual 
information, after the quench. This allowed us to 
prove a conjecture presented recently in Ref.~\cite{carollo2021emergent} for the case of the Ising chain.  

Our work opens several interesting research directions. In this paper we considered 
fermionic Hamiltonians and Lindblad operators. 
An interesting direction is to investigate 
whether the hydrodynamic framework can be extended to  spin degrees of freedom, for which 
the presence of the Jordan-Wigner string is expected to play an important role. 
Moreover, it would be interesting to consider localized dissipation, as for intstance 
done in Ref.~\cite{alba2021noninteracting}, Ref.~\cite{alba2021unbounded} or the 
combination of localized dissipation and driving~\cite{turkeshi2021diffusion,yamanaka2021exact}. 
One important direction is to try to extend the hydrodyanamic framework to interacting 
integrable systems. Recent years witnessed encouraging progress in this direction~\cite{deleeuw2021constructing,hutsalyuk2021integrability,medvedyeva2016exact,essler2020integrability,ziolkowska2020yang,bastianello2020generalized,friedman2020diffusive,vernier2020mixing,popkov2020inhomogeneous,popkov2020exact,bouchoule2020the,buca2020dissipative,popkov2021full}. It would be interesting to understand whether the structure 
of~\eqref{eq:f-entropy} remains the same for interacting integrable systems. 
Another possibility in order to assess the effect of interactions could be to use 
bosonization~\cite{bacsi2020vaporization}. 
A very promising direction is to extend our results to quenches from inhomogeneous 
initial states. It would be useful to understand whether the  approach of Ref.~\cite{bertini2018entanglement} and Ref.~\cite{alba2019entanglement} can be generalized in the presence of 
dissipation. Finally, it would interesting to investigate the effects of 
dissipation in the dynamics of entanglement and of quantum correlations in cellular automaton models, such as 
the rule 54 chain~\cite{klobas2021exact}. 

\section{Acknowledgements} 

F.C.~acknowledges support from the “Wissenschaftler-R\"uckkehrprogramm GSO/CZS” of the Carl-Zeiss-Stiftung and the German Scholars Organization e.V., as well as through the Deutsche Forschungsgemeinsschaft (DFG, German Research Foundation) under Project No. 435696605. V.A.~acknowledges support from the European Research Council under ERC Advanced grant No. 743032 DYNAMINT.

\appendix

\section{Subsystem entropies from the covariance matrix}
\label{sec:obs}

Entropy-related quantities and their dynamics can be obtained from the correlator $\Gamma$ 
(cf.~\eqref{eq:gamma-def}) (see Ref.~\cite{peschel2009reduced}). 
The single-block reduced density matrix $\rho_A$ can be written as 
\begin{equation}
	\label{eq:rho-majo}
	\rho_A=\frac{1}{2^\ell}\sum_{\{\mu\}}\mathrm{Tr}\big[\rho w_1^{\mu_1}w_2^{\mu_2}\cdots
	w_{2\ell}^{\mu_{2\ell}}\big] w_{2\ell}^{\mu_{2\ell}}\cdots w_1^{\mu_1}. 
\end{equation}
Here $w_i$ are Majorana fermions (cf.~\eqref{eq:majo}), 
$\mu_j=0,1$, and $\rho$ is the full-system density matrix. 
The Majorana correlation matrix $\Gamma_{ij}$ is defined as 
(cf.~\eqref{eq:gamma-def})
\begin{equation}
	\Gamma_{ij}:=\mathrm{Tr}(\rho w_i w_j)-\delta_{ij}. 
\end{equation}
Since Wick's theorem applies, the reduced density matrix can be recast in 
the form 
\begin{equation}
	\rho_A=\frac{1}{Z}e^{\frac{1}{4}\sum_{mn}w_m W w_n}, 
\end{equation}
where $Z$ ensures the normalization condition $\mathrm{Tr}(\rho_A)=1$. 
Here $W$ is related to $\Gamma$ as 
\begin{equation}
	\tanh\Big(\frac{W}{2}\Big)=\Gamma_\ell,  
\end{equation}
where  $\Gamma_\ell$ is obtained from $\Gamma_{ij}$ by restricting $i,j\in[1,2\ell]$.

First, by definition the $2\ell\times2\ell$ matrix $\Gamma$ is purely imaginary and 
antisymmetric and its eigenvalues are organized in pairs $\pm\nu_j$ with $j=1,\dots,\ell$. 
The R\'enyi entropies $S^{\scriptscriptstyle (n)}$ are written as 
\begin{equation}
	\label{eq:renyi-nu}
	S_A^{(n)}=\frac{1}{1-n}\sum_{j=1}^{\ell}\ln\Big[\Big(\frac{1+\nu_j}{2}\Big)^n+
	\Big(\frac{1-\nu_j}{2}\Big)^n\Big]. 
\end{equation}
Note that the sum in~\eqref{eq:renyi-nu} is restricted only to half of the eigenvalues 
of $\Gamma$, for instance the positive ones. 
The von Neumann entropy is obtained as 
\begin{equation}
	S_A=-\sum_{j=1}^{\ell}\frac{1+\nu_j}{2}\ln\Big(\frac{1+\nu_j}{2}\Big)
	+\frac{1-\nu_j}{2}\ln\Big(\frac{1-\nu_j}{2}\Big). 
\end{equation}
%

\section{A useful identity for the moments of the symbol of Majorana correlators}
\label{sec:useful-1}

Here we provide a useful identiy for the generic $2\times 2$ matrix $\mathbb{M}$ of 
the form 
\begin{equation}
	\label{eq:m-matrix}
	\mathbb{M}(a_j):=C\mathds{1}_2+A\sigma_x+B\sigma_y e^{i a_j\sigma_x}. 
\end{equation}
Here $\mathds{1}_2$ is the $2\times 2$ identity matrix, $\sigma_\alpha$ with 
$\alpha=x,y,z$ the standard Pauli matrices, $A,B,C\in\mathbb{C}$ arbitrary 
complex constants, and $a_j\in\mathbb{R}$ real parameters. Notice that the 
symbol~\eqref{eq:symbolo} of the generic Majorana correlation function is of the 
form~\eqref{eq:m-matrix} after a momentum-independent rotation $U=e^{i\pi\sigma_z/4}$. 
As this rotation is irrelevant for the calculation of the entropy, we are going 
to neglect it in the following. Let us consider the generic product 
\begin{equation}
	\label{eq:pin}
	\Pi_{n}(\{a_i\}):=\prod_{j=1}^{n}\mathbb{M}(a_j),\quad n\in\mathbb{N}. 
\end{equation}
For instance, $\Pi_{2}$ is given as  
\begin{multline}
	\label{eq:pi2}
	\Pi_{2}=\mathbb{M}(a_1)\mathbb{M}(a_2)=(C^2+A^2)\mathds{1}_2+2CA\sigma_x\\
	+CB\sigma_y(e^{ia_1\sigma_x}+e^{ia_2\sigma_x}) 
	+B^2e^{i(a_2-a_1)\sigma_x}+iAB\sigma_z(e^{ia_2\sigma_x}-e^{ia_1\sigma_x}). 
\end{multline}
For generic $n$, $\Pi_n$ will contain terms proportional to 
$\mathds{1}_2$ and to the Pauli matrices. Upon expanding the product in~\eqref{eq:pin}, 
one obtains the string of operators as 
\begin{equation}
	\label{eq:sigma-string}
	\Sigma=\Sigma_1\Sigma_2\quad\cdots\Sigma_n,\quad\mathrm{with}\, \Sigma_j=\mathds{1}_2,\sigma_x,\sigma_ye^{ia_j\sigma_x}.  
\end{equation}
Let us first consider the situation in which there are terms $\Sigma_{j_l}=\sigma_ye^{ia_{j_l}\sigma_x}$, 
with $1\le j_1<j_2<\cdots<j_l\in[1,n]$, and $l\in[0,n]$ is the total number of $\sigma_y$ terms 
present in the string. Notice that if $l=0$ the string is proportional either to $\mathds{1}_2$ or 
to $\sigma_x$. Let us also define as $z_k$ with $k\in[1,l+1]$ the number of $\sigma_x$ present between $j_k$ and $j_{k-1}$. 
Notice that $z_1$ is the number of $\sigma_x$ occurring at positions 
$k<j_1$ and $z_{l+1}$ at positions $k>j_l$. 
Now one can imagine of shifting all the terms $e^{i a_{j_l}\sigma_x}$ in $\Sigma$ 
to the right starting from the righmost one. In doing that one can use that 
\begin{equation}
	e^{ia_{j_l}\sigma_x}\sigma_y=\sigma_y e^{-ia_{j_l}\sigma_x}. 
\end{equation}
This allows to rewrite $\Sigma$ as 
\begin{equation}
	\label{eq:sigmap}
	\Sigma=\Sigma'e^{i\sum_{k=1}^l (-1)^k a_{j_k}}. 
\end{equation}
The string $\Sigma'$ is of the form 
\begin{equation}
	\label{eq:s1}
	\Sigma'=\sigma_x^{z_1}\mathds{1}_2^{j_1-1-z_1}\sigma_y\sigma_x^{z_2}
	\mathds{1}_2^{j_2-j_1-1-z_2}\sigma_y\cdots \sigma_y\sigma_x^{z_{l+1}}\mathds{1}_2^{n-j_l-z_{l+1}}.  
\end{equation}
By multiplying the string of operators  in~\eqref{eq:s1} one obtains 
\begin{equation}
	\Sigma'=(-1)^{\sum_k k z_k}\sigma_{z,l}
\end{equation}
with 
\begin{equation}
	\label{eq:szp}
	\sigma_{z,l}:=\left\{
	\begin{array}{cc}
		\sigma_y & z\,\mathrm{even}, l\,\mathrm{odd}\\
		-\sigma_x & z\,\mathrm{odd}, l\,\mathrm{even}\\
		-i\sigma_z & z=l\,\mathrm{odd}\\
		\mathds{1}_2 & z=l\,\mathrm{even}
	\end{array}
\right.
\end{equation}
where $l$ is the total number of $\sigma_y$ and $z=\sum_k z_k$. Notice 
that the operator $\sigma_{z,l}$ that one obtains by contracting the string 
depends only on $z,l$ and not on the ordering of the operators in the string. 
We now observe that each term with fixed position $j_1,\dots,j_l$ of 
$\sigma_y$ and a total number $z$ of $\sigma_x$ comes with multiplicity 
${\mathcal D}_z(\{j_l\})$, which is obtained by summing over all the ways 
of distributing the $\sigma_x$ and the identity matrix $\mathds{1}_2$. 
$\mathcal{D}_z$ is given as 
\begin{equation}
	\label{eq:mult0}
	{\mathcal D}_z(\{j_l\}):= 
	\sum_{z_1,z_2,\dots,z_{l+1}}(-1)^{\sum_m mz_m}B^{j_1,0}_{z_1}B^{j_{2,1}}_{z_2}\cdots
B^{j_{l,l-1}}_{z_l}B^{n-j_l}_{z_{l+1}}\delta_{z,\sum_m z_m},
\end{equation}
where we defined 
\begin{equation}
	B^x_y:=\binom{x}{y},\quad j_{\alpha,\beta}:=j_\alpha-j_\beta-1,\quad j_{1,0}:=j_1-1. 
\end{equation}
We do not provide the proof of~\eqref{eq:mult0}, which can be done by induction. We verified numerically  
for several values of $z$ and $l$ that~\eqref{eq:mult0} holds true. 

Putting everything together, we obtain that 
\begin{multline}
	\label{eq:exp}
	\Pi_{n}(\{a_i\})=\frac{1}{2}[(A-C)^n+(A+C)^n]\mathds{1}_2+
\frac{1}{2}[(A+C)^n-(A-C)^n]\sigma_x\\+
\sum_{z=0}^n\sum_{l=1}^{n-z}\sigma_{z,l}A^z B^l C^{n-l-z}\sum_{1\le j_1<j_2<\cdots<j_l\le n}
{\mathcal D}_z(\{j_l\})
\exp\Big(i\sum_{m=1}^l(-1)^{l-m}a_{j_m}\sigma_x\Big),
\end{multline}
where $\sigma_{s,l}$ is defined in~\eqref{eq:szp} and ${\mathcal D}_z$ in~\eqref{eq:mult0}. 
The first two terms in~\eqref{eq:exp} arise from contracting the strings of operators $\Sigma$ (cf.~\eqref{eq:sigma-string}) 
that do not contain any term $e^{ia_{j_l}\sigma_x}$. 

It is useful to consider the situation in which the exponent in the last term in~\eqref{eq:exp} 
depends only on the total number $l$ of $\sigma_y$ but not on the order in which they are 
placed. This will be relevant in~\ref{sec:useful-2}. We also restrict ourselves to even $n$. Thus, 
one can replace $j_l\to l$ in~\eqref{eq:exp} and perform the sum over $j_l$. 
First, one observes that for both $l,z$ odd the sum vanishes. In the other cases, one can 
verify that for any fixed $z$ and $l>0$ one has that 
\begin{equation}
	\label{eq:deg}
	\sum_{1\le j_1<j_2\cdots<j_l\le n} {\mathcal D}_z(\{j_l\})= 
	(-1)^{z}\binom{\lfloor (l+z)/2\rfloor}{\lfloor l/2\rfloor}\binom{n}{l+z}
\end{equation}
In summary, one obtains that~\eqref{eq:exp} is rewritten as 
\begin{multline}
	\label{eq:exp-1}
	\Pi_{n}(\{a_j\})=\frac{1}{2}[(A-C)^n+(A+C)^n]\mathds{1}_2+
\frac{1}{2}[(A+C)^n-(A-C)^n]\sigma_x\\+
\sum_{z=0}^n\sum_{l=1}^{n-z}\sigma'_{z,l}A^z B^l C^{n-z-l}
\binom{\lfloor (l+z)/2\rfloor}{\lfloor l/2\rfloor}\binom{n}{l+z}\exp\Big(i\sum_{m=1}^l(-1)^{l-m}a_{m}\sigma_x\Big), 
\end{multline}
where $\sigma'_{z,l}$ is zero if both $z,l$ are odd, and equal to $\sigma_{z,l}$ 
(cf.~\eqref{eq:szp}) otherwise. 
Equation~\eqref{eq:exp} allows us to calculate $\mathrm{Tr}(\Pi_{2n}(a(k)))$.
Only the terms with both $l$ and $z$ even survive 
in the last term in~\eqref{eq:exp}. We obtain that 
\begin{equation}
	\label{eq:scl}
	\mathrm{Tr}(\Pi_{2n})=(A-C)^{2n}+(A+C)^{2n}\\+
	2\sum_{z=0}^n\sum_{l=1}^{n-z}A^{2z} B^{2l} C^{2(n-z-l)}
	\binom{l+z}{l}\binom{2n}{2(l+z)}. 
\end{equation}
This is rewritten as 
\begin{equation}
	\label{eq:tr-full}
	\mathrm{Tr}(\Pi_{2n})=2\sum_{p=0}^{n}\binom{2n}{2p}(A^2+B^2)^p C^{2(n-p)}. 
\end{equation}
Notice that if $C=0$, which is the case considered in Ref.~\cite{calabrese2012quantum}, 
one obtains the simpler result 
\begin{equation}
	\mathrm{Tr}(\Pi_{2n})=2(A^2+B^2)^n. 
\end{equation}
%

\section{Hydrodynamic limit for the integer moments of $\Gamma$: Proof of a general formula}
\label{sec:useful-2}

In this section we derive a general formula describing the dynamics of the moments 
of the Majorana covariance matrix $\Gamma_\ell$ (see~\ref{sec:obs}) restricted 
to subystem $A$. We consider the case in which the full-system correlator $\Gamma$ is 
obtained from a symbol $\hat\Gamma_k$ 
of the form~\eqref{eq:symbolo}, i.e.,  
\begin{equation}
	\label{eq:tk}
	\hat\Gamma_k=C_k\mathds{1}_2+A_k\sigma^{(k)}_x+B_k\sigma^{(k)}_y e^{i a(k)t\sigma^{(k)}_x}. 
\end{equation}
Here $A_k,B_k,C_k$ are complex functions of $k$, $a(k)$ is a real function, and $k$ is 
the quasimomentum. Notice that here we are interested in the case with 
$C_k$ odd function of $k$, whereas $B_k,A_k$ are even functions of $k$ (see section~\ref{sec:K-par}). 
The fact that $C_k$  is odd ensures that $\mathrm{Tr}(\Gamma)=0$. However, 
for the derivation below the functions $A_k,B_k,C_k$ can be generic. 
In~\eqref{eq:tk} we introduced the rotated Pauli matrices $\sigma_\alpha^{(k)}$, 
which are defined as 
\begin{equation}
	\sigma_\alpha^{(k)}:=e^{i\vec v(k)\cdot\vec\sigma}\sigma_\alpha e^{-i\vec v(k)\cdot\vec{\sigma}},\quad \alpha=x,y,z. 
\end{equation}
Here $\vec v(k):=(v_x,v_y,v_y)$ is a vector of arbitrary real functions of $k$. We anticipate 
that the final result will not depend on the choice of $\vec v$. 
In the following, 
to lighten the notation, we are going to omit the dependence on $k$ in $A_k,B_k,C_k$. 
Let us now consider the correlation matrix $\Gamma_{rs}$, which is obtained as 
\begin{equation}
	\label{eq:g-def}
	\Gamma_{rs}=\int_{\pi}^\pi\frac{dk}{2\pi} e^{ik(r-s)} \hat\Gamma_k. 
\end{equation}
The restricted matrix $\Gamma_\ell$ is obtained by considering $r,s\in[1,\ell]$. 
Here we are interested in the dynamics of the moments of $\Gamma_\ell$, which are defined as 
\begin{equation}
	M_{2n}:=\mathrm{Tr}(\Gamma^{2n}_\ell). 
\end{equation}
Notice that we only consider the even moments of $\Gamma_\ell$ because the odd ones are zero 
by definition. Here we focus on the space-time scaling limit with $t,\ell\to\infty$ with 
their ratio fixed. The derivation that we are going to discuss is quite similar to Ref.~\cite{calabrese2012quantum}. 
To proceed we use the trivial identity 
\begin{equation}
	\label{eq:id}
	\sum_{m=1}^\ell e^{imk}=\frac{\ell}{4}\int_{-1}^1d\xi w(k)e^{i(\ell\xi+\ell+1)k/2},\quad w(k):=\frac{k}{\sin(k/2)}. 
\end{equation}
From~\eqref{eq:step1}, we obtain that 
\begin{equation}
	\label{eq:step1}
	\mathrm{Tr}(\Gamma^{2n}_\ell)=\Big(\frac{\ell}{4}\Big)^{2n}\int\limits_{[-\pi,\pi]^{2n}}\frac{d^{2n}k}{(2\pi)^{2n}}
	\int\limits_{[-1,1]^{2n}}d^{2n}\xi D(\{k\})F(\{k\})e^{i\ell\sum_{j=0}^{2n-1}\xi_j(k_{j+1}-k_j)/2}, 
\end{equation}
where we introduced the functions %
\begin{align}
	\label{eq:CC}
	D(\{k\})&=\prod_{j=0}^{2n-1}w(k_{j}-k_{j-1})\\
	\label{eq:F}
	F(\{k\})&=\mathrm{Tr}\prod_{j=0}^{2n-1}\hat \Gamma_{k_j}. 
\end{align}
Following Ref.~\cite{calabrese2012quantum}, it is convenient to change variables 
as
\begin{align}
&\zeta_0=\xi_1\\
&\zeta_i=\xi_{i+1}-\xi_i,\quad i\in[1,2n-1]. 
\end{align}
This allows us to rewrite~\eqref{eq:step1} as 
\begin{equation}
	\label{eq:tr}
	\mathrm{Tr}(\Gamma^{2n}_\ell)=\Big(\frac{\ell}{4}\Big)^{2n}\int\limits_{[-\pi,\pi]^{2n}}\frac{d^{2n}k}{(2\pi)^{2n}}
	\int_{R_\xi} d^{2n}\zeta_i D(\{k\})F(\{k\})e^{-i\ell\sum_{j=1}^{2n-1}\zeta_j(k_{j}-k_{0})/2}, 
\end{equation}
Here the integration domain for $\zeta_i$ is 
\begin{equation}
	R_\xi:-1\le\sum_{j=0}^{p-1}\zeta_j\le1, \quad p\in[1,2n]. 
\end{equation}
The integration over $\zeta_0$ is trivial because the integrand in\eqref{eq:tr} does not 
depend on $\zeta_0$. We obtain 
\begin{equation}
	\label{eq:tr}
	\mathrm{Tr}(\Gamma^{2n}_\ell)=\Big(\frac{\ell}{4}\Big)^{2n}\int\limits_{[-\pi,\pi]^{2n}}\frac{d^{2n}k}{(2\pi)^{2n}}
	\int d^{2n-1}\zeta_i D(\{k\})F(\{k\})e^{-i\ell\sum_{j=1}^{2n-1}\zeta_j(k_{j}-k_{0})/2}\mu(\{\zeta\}), 
\end{equation}
where we introduced the integration measure $\mu$ as 
\begin{equation}
	\label{eq:measure}
	\mu(\{\zeta_j\})=\max\Big[0,\min\limits_{j\in[0,2n-1]}
	\Big[1-\sum_{k=1}^j\zeta_k\Big]+\min\limits_{j\in[0,2n-1]}\Big[1+\sum_{k=1}^j\zeta_k\Big]\Big]. 
\end{equation}
The strategy to determine the behaviour of~\eqref{eq:tr}  in the space-time scaling limit is to 
use the stationary phase approximation for the integrals over $k_1,\dots,k_{2n-1}$ and $\zeta_1,\dots,\zeta_{2n-1}$. 
Stationarity with respect to the variables $\zeta_i$ in~\eqref{eq:tr} implies that 
\begin{equation}
	k_j\approx k_{0},\quad \forall j\in[1,2n-1]. 
\end{equation}
Now we can replace $k_j\to k_0$ in the definitions~\eqref{eq:CC} and~\eqref{eq:F} 
to obtain 
\begin{align}
	\label{eq:F-1}
	& F(\{k_j\})\to\mathrm{Tr}\prod_{j=0}^{2n-1}\Big[C_{k_0}\mathds{1}_2+A_{k_0}\sigma_x+B_{k_0}\sigma_ye^{i a(k_j)t\sigma_x}\Big]\\
	& D(\{k_j\})\to 2^n. 
\end{align}
Notice that in~\eqref{eq:F-1} we are not allowed to replace $k_j\to k_0$ in the phase 
factor $e^{ia(k_j)t\sigma_x}$, which has to be treated with the stationary phase. 
In~\eqref{eq:F-1} we replaced $\sigma_\alpha^{(k)}\to\sigma_\alpha$. We now use~\eqref{eq:exp}, which 
allows us to rewrite the product in~\eqref{eq:F-1}. From~\eqref{eq:tr} we obtain that 
\begin{multline}
	\label{eq:tr-2}
	\mathrm{Tr}(\Gamma^{2n}_\ell)=\Big(\frac{\ell}{2}\Big)^{2n}\int\limits_{[-\pi,\pi]^{2n}}\frac{d^{2n}k}{(2\pi)^{2n}}
	\int d^{2n-1}\zeta_i \mu(\{\zeta_m\})\mathrm{Tr}\Big\{
		\frac{1}{2}[(A-C)^{2n}+(A+C)^{2n}]\mathds{1}_2\\+
		\frac{1}{2}[(A+C)^{2n}-(A-C)^{2n}]\sigma_x+
		\sum_{z=0}^{2n}\sum_{l=1}^{2n-z}\sigma_{z,l}A^z B^l C^{2n-z-l}\\
		\times\sum_{1\le j_1<j_2<\cdots<j_l\le 2n}\Big[
{\mathcal D}_z(\{j_m\})
\exp\Big(i\sum_{m=1}^l(-1)^{l-m}a(k_{j_m})t\sigma_x\Big)
	\Big]\Big\}e^{-i\ell\sum_{j=1}^{2n-1}\zeta_j(k_{j}-k_{0})/2}, 
\end{multline}
where ${\mathcal D}_z$ and $\sigma_{z,l}$ are defined in~\eqref{eq:mult0}
It is straightforward to perform the trace of the first two terms in the curly brackets. In the last term in~\eqref{eq:tr-2} only 
the cases with both $l$ and $z$ even give a nonzero contribution. Moreover, the last integral in~\eqref{eq:tr-2} 
is invariant under permutation of the momenta $k_j$. Thus, one can replace $a(k_{j_m})\to a(k_m)$ in the exponential. 
After using~\eqref{eq:deg} and performing the trace we obtain 
\begin{multline}
	\label{eq:tr-3}
	\mathrm{Tr}(\Gamma^{2n}_\ell)=\Big(\frac{\ell}{2}\Big)^{2n}\int\limits_{[-\pi,\pi]^{2n}}\frac{d^{2n}k}{(2\pi)^{2n}}
	\int d^{2n-1}\zeta_i\mu(\{\zeta_m\})
e^{-i\ell\sum_{j=1}^{2n-1}\zeta_j(k_{j}-k_{0})/2}	
\Big\{\\
[(A-C)^{2n}+(A+C)^{2n}]+
2\sum_{z=0}^{n}\sum_{l=1}^{n-z}
\binom{l+z}{l}\binom{2n}{2(l+z)}
A^{2z} B^{2l} C^{2(n-l-z)}\cos\Big(\sum_{m=0}^{2l+1}(-1)^{2l-m+1}a(k_m)t\Big)
\\+
2i\sum_{z=0}^{n-1}\sum_{l=1}^{n-z-1}
\binom{l+z}{l}\binom{2n}{2l+2z+1}
A^{2z+1} B^{2l} C^{2n-2z-2l-1}\sin\Big(\sum_{m=0}^{2l-1}(-1)^{2l-m+1}a(k_m)t\Big)
\Big\}, 
\end{multline}
To proceed, we employ the stationary phase approximation to extract the 
leading behavior of~\eqref{eq:tr-3} in the limit $\ell,t\to\infty$ with 
the ratio $t/\ell$ fixed. 
By rewriting the sine and cosine function in~\eqref{eq:tr-3} in terms of 
exponentials, it is clear that 
one has integrals $\Lambda_l$ of the type 
\begin{multline}
	\label{eq:lambda-l}
\Lambda_l:=\Big(\frac{\ell}{2}\Big)^{2n-1}
\int\limits_{[-\pi,\pi]^{2n-1}}\frac{d^{2n-1}k}{(2\pi)^{2n-1}}
\int d^{2n-1}\zeta_i f(k_0)\\	
\exp\Big(\pm i\sum_{j=0}^{2l-1}(-1)^{2l-j-1}a(k_j)t
-i\ell\sum_{j=1}^{2n-1}\zeta_j(k_{j}-k_{0})/2\Big)\mu(\{\zeta_m\}), 
\end{multline}
where $f(k_0)$ is obtained from~\eqref{eq:tr-3} by collecting the terms 
that do not contain complex exponentials. The subscript $l$ in 
$\Lambda_l$ is to stress that $l$ appears in the exponent in~\eqref{eq:tr-3} and 
it affects the stationary phase result. The first term in the exponential 
in~\eqref{eq:lambda-l} is obtained from the cosine and sine functions in~\eqref{eq:tr-3}, 
whereas the second one is the phase factor in~\eqref{eq:tr-3}. 

We are now ready to apply the stationary phase approximation to the 
integral~\eqref{eq:lambda-l}. 
The stationary phase states that in the limit $\ell\to\infty$ one has~\cite{wong}  
\begin{equation}
	\label{eq:s-phase}
	\int_D d^N xp(\vec x)e^{i\ell q(\vec x)}
	\rightarrow\Big(\frac{2\pi}{\ell}\Big)^{N/2}p(\vec x_0)|\mathrm{det}H|^{-1/2}\exp\Big[i\ell q(\vec x_0)+i\pi\frac{\sigma_A}{4}\Big]. 
\end{equation}
Here $p(\vec x)$ and $q(\vec x)$ are arbitrary functions, $D$ is the integration domain and 
$\ell$ is a parameter. On the right hand side in~\eqref{eq:s-phase}, 
$\vec x_0$ is the stationary point satisfying 
$\vec\nabla q(\vec x_0)=0$. In~\eqref{eq:s-phase} $H$ is the 
hessian matrix $H=\partial_{x_i}\partial_{x_j} q(\vec x)$, 
and $\sigma$ its signature, i.e., the difference between the number of positive and negative 
eigenvalues of $H$. 

From~\eqref{eq:lambda-l}, in the hydrodynamic limit $\ell,t\to\infty$ with 
their ratio fixed, the stationary phase approximation in 
the variables $k_1,\dots, k_{2n-1}$ 
and $\zeta_1,\dots,\zeta_{2n-1}$ is determined by the stationary points 
\begin{align}
&\bar k_j=k_0 & & j=1,\dots,2n-1\\
\label{eq:zeta-sign}
&\bar \zeta_j=\pm 2\frac{t}{\ell}(-1)^j a'(k_j) & & j=1,\dots, 2l-1\\
&\bar \zeta_j=0 &&j=2l,\dots,2n-1,
\end{align}
where the $\pm$ in~\eqref{eq:zeta-sign} originates from the first 
term in the exponent in~\eqref{eq:lambda-l}. 
From~\eqref{eq:measure}, one obtains thata at the stationary point 
\begin{equation}
	\label{eq:mu-sp}
	\bar\mu(\{\bar\zeta_j\})=\left\{
		\begin{array}{cc}
			\frac{2}{\ell}\max[0,\ell-|a'(k_0)|t] & l\ne0\\
			2 & l=0
		\end{array}
	\right.
\end{equation}
Importantly, Eq.~\eqref{eq:mu-sp} does not depend on the $\pm$ 
sign of $\bar \zeta_j$ in~\eqref{eq:zeta-sign}. 

To proceed, we observe that in our case 
$\mathrm{det}(H)=4^{1-2n}$ (cf.~\eqref{eq:s-phase}). 
Moreover, the signature $\sigma$ is always zero, and the 
phase factor~\eqref{eq:s-phase} does not contribute because the alternating sum in the 
exponent in~\eqref{eq:s-phase} has an even number of terms. 
Crucially, since $\bar\mu(\{\bar\zeta_m\})$ does 
not depend on the sign in~\eqref{eq:zeta-sign}, the last term in~\eqref{eq:tr-3} vanishes. 
Finally, we obtain that 
\begin{multline}
	\label{eq:tr-sf}
	\mathrm{Tr}(\Gamma^{2n}_\ell)=\ell\int\frac{dk}{2\pi}[(A-C)^{2n}+(A+C)^{2n}]\\
	+2\sum_{z=0}^n\sum_{l=1}^{n-z}\binom{l+z}{l}\binom{2n}{2(l+z)}
	\int\frac{dk}{2\pi} A^{2z}B^{2l}C^{2(n-l-z)}\max(0,\ell-|a'|t). 
\end{multline}
Here we redefined $k_0\to k$. 
For $C=0$ one recovers the result of Ref.~\cite{calabrese2012quantum}. 
Eq.~\eqref{eq:tr-sf} can be conveniently rewritten as 
\begin{multline}
	\label{eq:tr-fin-1}
	\mathrm{Tr}(\Gamma^{2n}_\ell)=\int\frac{dk}{2\pi}[(A-C)^{2n}+(A+C)^{2n}]\min(\ell,|a'|t)\\
	+2\sum_{z=0}^n\binom{2n}{2z}
	\int\frac{dk}{2\pi} (A^{2}+B^{2})^zC^{2(n-z)}\max(0,\ell-|a'|t). 
\end{multline}
Equivalently, one can rewrite~\eqref{eq:tr-fin-1} as 
\begin{multline}
	\label{eq:tr-fin}
	\mathrm{Tr}(\Gamma^{2n}_\ell)=\int\frac{dk}{2\pi}[(A-C)^{2n}+(A+C)^{2n}]\min(\ell,|a'|t)\\
	+\int\frac{dk}{2\pi}\mathrm{Tr}(\Gamma_L^{2n})\max(0,\ell-|a'|t). 
\end{multline}
In the second row in~\eqref{eq:tr-fin} we used~\eqref{eq:tr-full} to identify 
the trace of the moments of the full-system correlator $\Gamma_L$. 
Eq.~\eqref{eq:tr-fin} implies that for a generic function ${\mathcal F}(z)$, one has 
that 
\begin{multline}
	\label{eq:tr-F-fin}
	\mathrm{Tr}({\mathcal F}(\Gamma^2_\ell))=\int\frac{dk}{2\pi}[{\mathcal F}((A-C)^{2})+{\mathcal F}((A+C)^{2})]\min(\ell,|a'|t)\\
	+\int\frac{dk}{2\pi}\mathrm{Tr}({\mathcal F}(\Gamma_L^{2}))\max(0,\ell-|a'|t). 
\end{multline}
Eq.~\eqref{eq:tr-F-fin} is equivalent to~\eqref{eq:tr-F-man} after replacing 
$a(k)=2\varepsilon_h(k)$ (cf.~\eqref{eq:vareps}). 

\begin{figure}[t]
\begin{center}
\includegraphics[width=.5\textwidth]{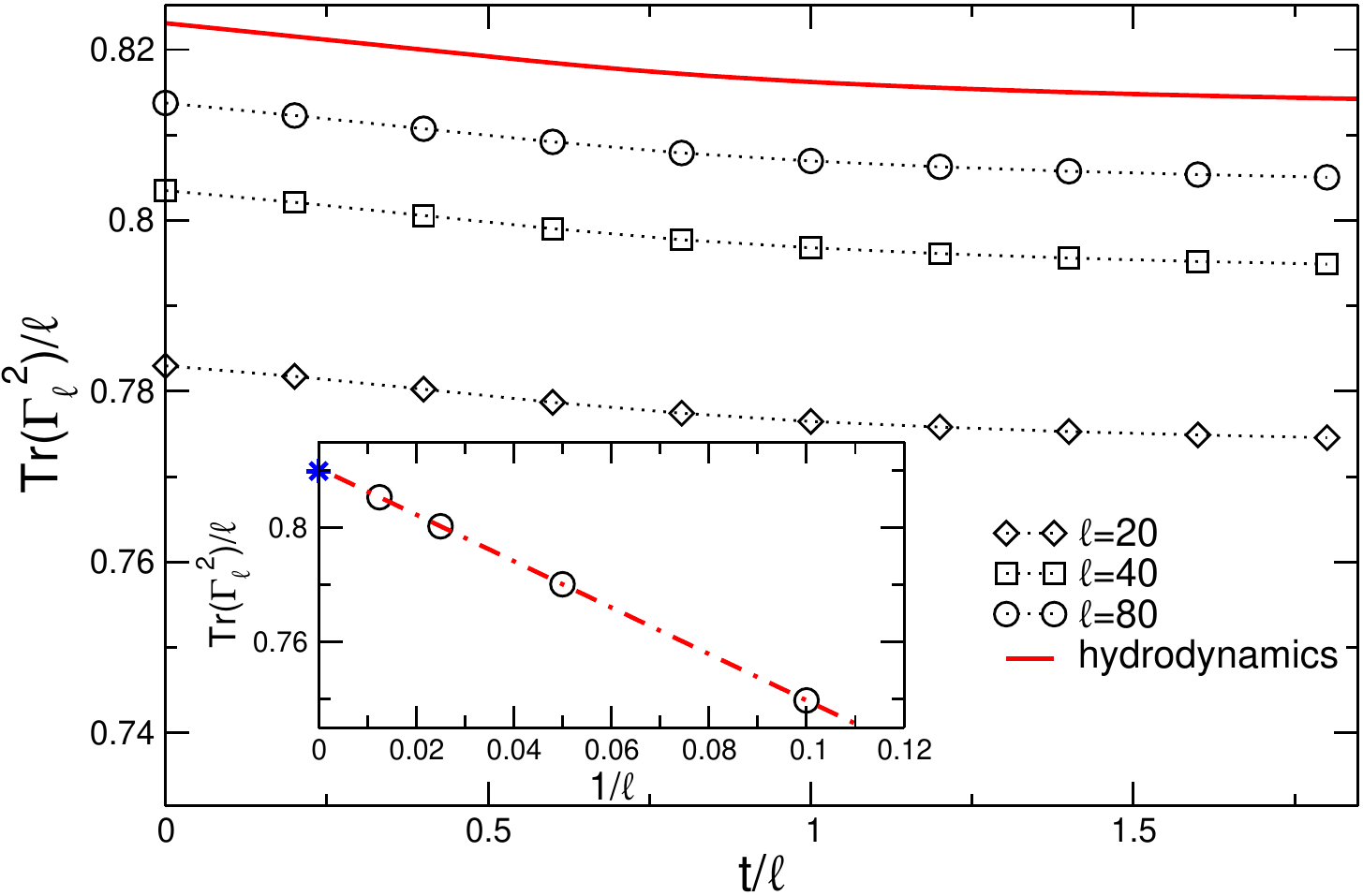}
\caption{ Numerical check of Eq.~\eqref{eq:tr-fin} for $n=1$. We plot 
 $\mathrm{Tr}(\Gamma_\ell^{2n})/\ell$ versus $t/\ell$. 
 The functions $A_k,B_k,C_k$ are defined in~\eqref{eq:A}-\eqref{eq:C}. 
 Here we choose $a(k)=2\cos(k)$. The symbols are exact lattice results 
 for different subsystem sizes $\ell$. The continuous red line is 
 the result in the scaling limit $\ell,t\to\infty$ with $t/\ell$ fixed. 
 Finite-size and finite-time corrections are visible. Scaling corrections 
 are investigated in the inset plotting $\mathrm{Tr}(\Gamma_\ell^{2n})/\ell$ 
 versus $1/\ell$ at fixed $t/\ell=0.4$. The star symbol is the result 
 in the hydrodynamic limit $\ell\to\infty$. The dashed-dotted line is 
 a linear fit. 
}
\label{fig8:check}
\end{center}
\end{figure}

It is important to check the validity of~\eqref{eq:tr-F-fin}. 
In Fig.~\ref{fig8:check} we discuss some numerical checks of~\eqref{eq:tr-F-fin} for 
the second moment $\mathrm{Tr}(\Gamma_\ell^2)$ of $\Gamma_\ell$. 
We consider the fermionic correlator $\Gamma$ of the form~\eqref{eq:symbolo} with 
\begin{align}
	\label{eq:A}
	& A_k=0.87\cos(k)\\
	\label{eq:B}
	& B_k=0.1235\cos^2(k)\\
	\label{eq:C}
	& C_k=0.234\sin(k). 
\end{align}
We also fix $a(k)=2\cos(k)$. 
The symbols in Fig.~\ref{fig8:check} are exact numerical data for finite 
$\ell$ and $t$. Since we are interested in the space-time scaling limit, 
in the figure we plot $\mathrm{Tr}(\Gamma_\ell^2)/\ell$ versus $t/\ell$. 
The continuous line in Fig.~\ref{fig8:check} is the result~\eqref{eq:tr-fin} 
for $n=1$. As it is clear from the figure the data exhibit strong finite-size 
and finite-time corrections. However, upon increasing $\ell$ they approach the 
analytic result~\eqref{eq:tr-fin}. A more systematic analysis of the 
scaling corrections is presented in the inset of Fig.~\ref{fig8:check}, 
showing $\mathrm{Tr}(\Gamma_\ell^2)/\ell$ versus $1/\ell$ at fixed 
$t/\ell=0.4$. The star symbol in the inset is the expected result in the 
hydrodynamic limit. The dashed-dotted line is a linear fit. The quality 
of the fit confirms the validity of~\eqref{eq:tr-fin} and suggests that 
the corrections are ${\mathcal O}(1/\ell)$. Similar corrections are 
observed also in the case without dissipation~\cite{alba2018entanglement}.

\section*{References}
\bibliographystyle{iopart-num.bst}
\bibliography{bibliography}

\end{document}